\documentclass[a4paper,12pt]{article}
\usepackage{epsfig}
\usepackage{amsfonts}

\newlength{\extraspace}
\newlength{\extraspaces}
\catcode`\@=11
 
\def\numberbysection{\@addtoreset{equation}{section}
\def\theequation{\arabic{section}.\arabic{equation}}}

\renewcommand{\theequation}{\arabic{section}.\arabic{equation}}
\begin{document}
%
\thispagestyle{empty}
\begin{center}
\begin{flushright}
TIT/HEP--445 \\
{\tt hep-th/0004188} \\
April, 2000 \\
\end{flushright}
\vspace{3mm}
\begin{center}
{\Large
{\bf Nonnormalizable Zero Modes on BPS 
Junctions 
 }} 
\\[18mm]

{\sc Kenji~Ito},\footnote{
\tt e-mail: kito@th.phys.titech.ac.jp} \hspace{2.0mm}
{\sc Masashi~Naganuma},\footnote{
\tt e-mail: naganuma@th.phys.titech.ac.jp} \hspace{2.0mm}
{\sc Hodaka~Oda},\footnote{
\tt e-mail: hoda@th.phys.titech.ac.jp} \hspace{2.0mm}
and \hspace{2.0mm}
{\sc Norisuke~Sakai}\footnote{
\tt e-mail: nsakai@th.phys.titech.ac.jp} \\[3mm]
{\it Department of Physics, 
Tokyo Institute of Technology \\[2mm]
Oh-okayama, Meguro, Tokyo 152-8551, Japan} \\[4mm]

\end{center}
\vspace{18mm}
{\bf Abstract}\\[5mm]
{\parbox{13cm}{\hspace{5mm}
Using an exact solution as a concrete example, Nambu-Goldstone 
modes on the BPS domain wall junction are worked out for ${\cal N}=1$ 
supersymmetric theories in four dimensions. 
Their wave functions extend along the wall to infinity (not localized) 
and are not normalizable. 
It is argued that this feature is a generic phenomenon of Nambu-Goldstone 
modes on domain wall junctions in the bulk flat space in any 
dimensions. 
We formulate mode equations and show that fermion and boson with 
the same mass come in pairs except massless modes which can appear singly, 
in accordance with unitary representations of $(1, 0)$ 
supersymmetry. 

}}
\end{center}
\vfill
\hspace*{10mm}
PACS number(s): 11.27.+d,11.30.P,11.15.Kc,11.10.-z 
\newpage
\vfill
\newpage
\setcounter{equation}{0}
\setcounter{footnote}{0}

\section{Introduction}

In recent years an interesting idea has been advocated to regard 
 our world as a domain wall embedded in higher dimensional spacetime
 \cite{ADD}. 
Most of the particles in the standard model should be realized as 
modes localized on the wall. 
Phenomenological implications of the idea have been extensively studied 
from many aspects. 
Another fascinating possibility has also been proposed 
to consider walls in the bulk spacetime which has negative cosmological 
constant \cite{RS}. 
The model can give large mass hierarchy or can give massless graviton 
localized on the wall. 
Subsequently a great deal of research activity has been performed 
to study and extend the proposal \cite{BehrndtCvetic}. 

Since walls typically have co-dimension one, it is desirable to consider 
intersections and/or junctions of walls in order to obtain our 
four dimensional world from a spacetime with much higher dimensions. 
The model with the bulk cosmological constant has been  extended 
to produce an intersection of walls \cite{ADDK}. 

Supersymmetry has been useful to achieve stability of solitonic solutions 
such as domain walls. 
Domain walls in supersymmetric theories can saturate the Bogomol'nyi 
bound \cite{BPS}. 
Such a domain wall preserves half of the original supersymmetry and is 
called a $1/2$ BPS state \cite{WittenOlive}. 
It has also been noted that these BPS states possess a topological charge 
which becomes a central charge $Z$ of the supersymmetry algebra 
\cite{AbrahamTownsend} \cite{DvaliShifman}. 
Thanks to the topological charge, these BPS states are guaranteed to be stable 
under arbitrary local fluctuations. 
Various properties of domain walls in ${\cal N}=1$ supersymmetric 
field theories in four dimensions have been extensively studied 
 \cite{KSS}, \cite{KSY}. 
In particular the modes on the domain wall background 
have been worked out 
and are found to contain fermions and/or bosons localized on the wall 
in many cases \cite{RubakovShaposhnikov}, \cite{ChibisovShifman}. 

Recently domain wall junctions have attracted 
much attention as another interesting possibility for BPS states 
\cite{GibbonsTownsend}--\cite{GorskyShifman}. 
Domain walls occur in interpolating two discrete degenerate vacua 
in separate region of space. 
If three or more different discrete vacua occur in separate 
region of space, segments of domain walls separate 
each pair of the neighboring vacua. 
If the two spatial dimensions of all of these domain walls have 
one dimension in common, 
these domain walls meet at a one-dimensional junction. 
The solitonic configuration for the junction can 
preserve a quarter of supersymmetry and is called a $1/4$ BPS state. 
There has been progress to study general properties of such domain wall 
junctions. 
{}For instance a new topological charge $Y$ is found to appear for such 
a $1/4$ BPS state \cite{GibbonsTownsend}--
\cite{GorskyShifman}. 
If we start from ${\cal N}=1$ four dimensional supersymmetric field theories, 
the domain wall junction preserves only one supercharge. 
Consequently the resulting theory was expected to be a $(1, 0)$ 
supersymmetric theory in $1 + 1$ dimensions \cite{GibbonsTownsend} 
which offers an intriguing possibility of chiral fermions. 
Moreover, there have been a number of numerical simulations 
which indicate the existence of the domain wall junction solutions 
 \cite{Saffin}. 
In spite of all these efforts, 
it has been difficult to obtain an explicit solution  and to prove the 
existence of a BPS domain wall junction. 

Recently we have succeeded to work out an exact solution for the 
BPS domain wall junction for the first time \cite{HKMN}. 
The exact solution allows a thorough study of the properties of the BPS 
domain wall junction. 
Consequently several misconceptions can be pointed out and rectified. 
One such point is the sign and meaning of the new central charge $Y$ 
which arises when walls form a junction. 
Our exact solution showed that the central charge $Y$ contributes negatively 
to the mass of the domain wall junction configuration. 
Therefore we should not consider the central charge $Y$ alone as the mass 
of the junction. 
Various other aspects of the domain wall junctions are also studied recently 
\cite{BinosiVeldhuis} -- \cite{NamOlsen}. 

The purpose of the present paper is to give a more detailed study of the 
properties of the BPS domain wall junction in ${\cal N}=1$ supersymmetric 
field theories. 
We study the modes on the 
background of the domain wall junction, 
especially the Nambu-Goldstone modes. 
We will use our exact solution as a concrete example and will extract 
the generic properties of the BPS domain wall junctions. 
We define mode equations and demonstrate explicitly that fermion and 
boson with the same mass have to come in pairs except massless modes. 
Massless modes can appear singly without accompanying fields with opposite 
statistics. 
We also show that unitary representations of 
the surviving $(1, 0)$ supersymmetry are classified into doublets for 
massive modes and singlets for massless modes. 
We work out 
explicitly massless Nambu-Goldstone modes associated 
with the broken supersymmetry and translational invariance. 
We find that the Nambu-Goldstone fermions exhibit an 
 interesting chiral structure in accordance with the surviving 
 $(1, 0)$ supersymmetry algebra. 
However, we also find that any linear combinations of the 
Nambu-Goldstone modes associated with the junctions become 
a linear combination of zero modes on at least one of the domain walls 
asymptotically along these walls. 
Since their wave functions are extended along these walls without 
damping, they are not localized states on the junction. 
Therefore they are not normalizable, contrary to a previous expectation 
\cite{GibbonsTownsend}. 
This indicates that the resulting theory cannot be regarded as a genuine 
$1+1$ dimensional field theory with discrete particle spectrum even at zero 
energy. 
Although the remaining supersymmetry is just $(1, 0)$ which is characteristic 
to $1+1$ dimensions, 
we have to keep in mind that the domain wall junction configuration 
is actually living in one more dimensions similarly to the domain wall itself. 
Zero modes on the junction turn out to have properties quite similar to 
those on the domain wall. 
The non-normalizability of Nambu-Goldstone modes on the 
junction configuration is not an accident in this particular model. 
We observe that the origin of this property can be traced back to the fact 
that the supersymmetry is broken by the coexistence of nonparallel walls. 
Therefore 
the fact that the Nambu-Goldstone modes on the BPS domain wall junction 
are not normalizable 
is a generic feature of supersymmetric field 
theories in the bulk flat space. 

One should note that our conclusion need not apply to the case with negative 
cosmological constant in the bulk. 
In the presence of a bulk negative cosmological constant in six dimensions, 
five dimensional walls can intersect in Anti de Sitter space. 
If one demands a flat space at the four dimensional intersection, one has 
an Anti de Sitter space not only in the bulk but also even on the walls
\cite{ADDK}. 
Since Anti de Sitter space does not have 
translational invariance, 
the wave function of the zero mode does not become constant along 
the wall asymptotically, contrary to our situation. 
If one approaches the intersection along the wall, one meets precisely 
the same situation as the wall in the five dimensional Anti de Sitter 
space. 
{}For instance graviton zero mode is exponentially suppressed away from the 
intersection along the wall direction to produce a normalizable wave function. 
Therefore the Anti de Sitter geometry along the wall plays an essential role 
to achieve the localization of the wave function on the intersection 
in models with cosmological constant.

In sect.~2, we introduce BPS equations and the exact solution for the 
domain wall junction and discuss representations of the surviving 
$(1, 0)$ supersymmetry algebra. 
In sect.~3, we present mode equations which define the fluctuations 
on the background of domain wall junction. 
We work out the Nambu-Goldstone mode explicitly and show that they 
are not normalizable. 
Physical origin of the nonnormalizability is clarified and the general 
validity of this phenomenon is argued. 
In sect.~4, the relation between the choice of BPS equations 
and the boundary condition is discussed for a general Wess-Zumino model. 
In sect.~5, the central charge density and the energy density 
are examined and an interesting behavior is observed. 
The fermionic contributions to the central charges and mode equations 
in a convenient gamma matrix representation are given in appendices.

\section{BPS equations and the $(1, 0)$ supersymmetry algebra }

\subsection{Two $1/4$ BPS states and two BPS equations}
It is known that  
if the translational invariance is broken as is the case for domain 
walls and/or junctions, the ${\cal N}=1$ superalgebra in general receives 
contributions from central charges 
\cite{Townsend1}, 
\cite{DvaliShifman}--\cite{HKMN}, \cite{Townsend2}. 
The anti-commutator between two left-handed supercharges has central 
charges 
$Z_k$,  $k=1,2,3$ 
\begin{equation}
\{ Q_\alpha, Q_\beta  \} =
2 i (\sigma^k \bar{\sigma}^0 )_\alpha{}^{\gamma} 
\epsilon_{\gamma \beta} Z_k. 
\label{qqantcom}
\end{equation}
Here and the following we use two-component spinors following the 
convention of ref.\cite{WessBagger} except that the four dimensional indices 
are denoted by Greek letters $\mu, \nu =0, 1, 2, 3$ instead of 
roman letters $m, n $. 
The anti-commutator between left- and right-handed supercharges 
receives a contribution from central charges $Y_k, \ k=1,2,3$  
besides the energy-momentum 
four-vector $P^{\mu}, \ {\mu}=0, \cdots, 3$ of the system 
\begin{equation}
\{ Q_\alpha, \bar{Q}_{\dot{\alpha}} \} =
2 (\sigma^{\mu}_{\alpha \dot{\alpha}} P_{\mu} 
+ \sigma^k_{\alpha \dot{\alpha}} Y_k ) .
\label{qqantcomy}
\end{equation}
One may call $Z_k$ and $Y_k$ as $(1, 0)$ and $(1/2, 1/2)$ central charges 
in accordance with the transformation properties under the Lorentz group. 
Central charges, $Z_k$ and $Y_k$, come from the total divergence, 
and they are non-vanishing when there are nontrivial differences 
in asymptotic behavior of fields in different region of spatial infinity 
as is the case of domain walls and junctions \cite{CHT}. 
Therefore these charges are topological in the sense that they are 
determined completely by the boundary conditions at infinity. 
{}For instance, we can 
take a general Wess-Zumino model with 
an arbitrary number of chiral superfields $\Phi^i$, 
an arbitrary superpotential ${\cal W}$ 
 and an arbitrary K\"ahler potential $K(\Phi^i, \Phi^{* j})$ 
\begin{equation}
{\cal L}=\int d^2\theta d^2\bar{\theta} K(\Phi^i,\Phi^{* j}) 
 + \left[ \int d^2\theta {\cal W}(\Phi^i) 
+ \mbox{h.c.} \right],
\label{generalWZmodel}
\end{equation}
and compute the anticommutators (\ref{qqantcom}), 
(\ref{qqantcomy}) to find the central charges. 
The contributions to these central charges from bosonic components of 
chiral superfields are given \footnote{The central charge $Y_k$
also receives contributions from fermionic components of chiral
superfields which is given in appendix A.}
by \cite{CHT}
\begin{equation}
Z_k = 2 \int d^3 x \, \partial_k {\cal W}^*(A^*),
\label{centralchargeZ}
\end{equation}
\begin{equation}
Y_k =i \epsilon^{knm}
\int d^3 x \, K_{i j^*}
\partial_n (A^{*j} \partial_m A^i), \qquad  \epsilon^{123}=1, 
\label{centralchargeY}
\end{equation}
where $A^i$ is the scalar component of the $i$-th chiral 
superfield $\Phi^i$ and  
$ K_{i j^*}=\partial^2 K(A^*, A) /\partial A^i\partial A^{*j}$ 
is the K\"ahler metric.

BPS domain wall is a $1/2$ BPS state \cite{DvaliShifman} 
and BPS domain wall junction is a $1/4$ BPS state
\cite{GibbonsTownsend}\cite{CHT}. 
To find the BPS equations satisfied by these BPS states, 
we consider a hermitian linear combination of operators 
$Q$ and $\bar{Q}$ with an arbitrary complex two-vector $\beta^\alpha$ 
and its complex conjugate $\bar{\beta}^{\dot{\alpha}} = (\beta^\alpha)^*$ 
as coefficients 
\begin{equation}
K= \beta^\alpha Q_\alpha + 
\bar{\beta}^{\dot{\alpha}} \bar{Q}_{\dot{\alpha}}. 
\end{equation}
We treat $\beta^\alpha$ as c-numbers rather than
the Grassmann numbers. 
Since $K$ is hermitian, 
the expectation value of the square of $K$ over any state 
is non-negative definite, $\langle S | K^2 | S \rangle \ge 0$.
The field configuration of static junction must be at least two-dimensional. 
If we assume, for simplicity, that 
it depends on $x^1$, $x^2$ 
then we obtain $\langle Z_3 \rangle=\langle Y_1 \rangle
=\langle Y_2 \rangle = 0$ from Eqs.(\ref{centralchargeZ}) 
and (\ref{centralchargeY}), and 
the inequality implies in this case 
\begin{eqnarray}
\langle H \rangle
&\!\!\! \ge &\!\!\!
\frac{-1}{|\beta^1|^2+|\beta^2|^2} \Biggl\{
(|\beta^1|^2-|\beta^2|^2) \langle Y_3 \rangle
+ {\rm Re} \left[(\beta^1)^2 \langle - Z_2 -i Z_1 \rangle \right]
\nonumber \\
&\!\!\!  &\!\!\! {}
+ {\rm Re} \left[(\beta^2)^2 \langle - Z_2 +i Z_1 \rangle \right]
\Biggr\},
\end{eqnarray}
for any $\beta^\alpha$ and for any state.  
The equality holds if and only if the linear combination of supercharges, 
$K$, is preserved by the state $ |S \rangle$
\begin{equation}
K \left\vert S \right\rangle= 0 . 
\end{equation}
In this case, the state $ |S \rangle$ saturates the energy bound and 
is called a BPS state. 
We find that there are two candidates 
for the saturation of the energy bound \cite{HKMN}; 
\begin{eqnarray}
H = H_{\rm I} \equiv
|\langle -i Z_1-Z_2 \rangle|
-\langle Y_3 \rangle, \  \ \mbox{when} \ \ 
\bar{\beta}^{\dot{1}}=\beta^1
\frac{\langle i Z_1+Z_2 \rangle}{{|\langle i Z_1+Z_2 \rangle|}}, \ \ 
\beta^2=\bar{\beta}^{\dot{2}}=0,  \\
H = H_{\rm II} \equiv
|\langle i Z_1-Z_2 \rangle|
+\langle Y_3 \rangle, \ \ \mbox{when} \ \ 
\beta^1=\bar{\beta}^{\dot{1}}=0, \ \
\bar{\beta}^{\dot{2}}=\beta^2
\frac{\langle -i Z_1+Z_2 \rangle}{{|\langle -i Z_1+Z_2 \rangle|}}.
\end{eqnarray}

In the case of $H_{\rm I}\ne H_{\rm II}$, 
the BPS bound becomes 
$
\langle H \rangle \ge 
{\rm max}\{ 
H_{\rm I} , H_{\rm II}
\} .
$
If $H_{\rm I} > H_{\rm II}$, 
then supersymmetry can only be preserved at $\langle H \rangle=H_{\rm I}$ 
and 
the only one combination of supercharges is  conserved 
\begin{equation}
\left(
Q_1 + 
\frac{\langle i Z_1+Z_2 \rangle}
{|\langle i Z_1+Z_2 \rangle|}
\bar{Q}_{\dot{1}}
\right)
| \langle H \rangle=H_{\rm I} \rangle =0.
\label{conseved_charge_1st}
\end{equation}
If $H_{\rm II}>H_{\rm I}$, 
then supersymmetry can only be preserved at 
$\langle H \rangle = H_{\rm II}$ and
the only one combination of supercharges is  conserved 
\begin{equation}
\left(
Q_2 + 
\frac{\langle -i Z_1+Z_2 \rangle}
{|\langle -i Z_1+Z_2 \rangle|}
\bar{Q}_{\dot{2}}
\right)
| \langle H \rangle=H_{\rm II} \rangle =0.
\label{conseved_charge_2st}
\end{equation}
In the case of $H_{\rm I}=H_{\rm II}$, 
two candidates of BPS bounds coincide 
 and BPS state conserves both of two supercharges, 
(\ref{conseved_charge_1st}) and (\ref{conseved_charge_2st}); 
this is a $1/2$ BPS state. 

{}For the general Wess-Zumino model in Eq.(\ref{generalWZmodel}), 
the condition of supercharge conservation (\ref{conseved_charge_1st}) 
for $H=H_{\rm I}$ applied to chiral superfield 
$\Phi^i=(A^i, \psi^i, F^i)$ gives 
after eliminating the auxiliary field $F^i$ 
\begin{equation}
2 
{\partial A^i \over \partial \bar{z}} 
=- \Omega_{+} F^i 
= \Omega_{+}K^{-1 i j^*} \frac{\partial {\cal W}^*}{ \partial A^{*j}}, \quad 
\Omega_{+}\equiv i \frac{\langle -i Z_1^*+Z_2^* \rangle}
{|\langle -i Z_1^*+Z_2^* \rangle|},
\label{Be1}
\end{equation}
where complex coordinates $z=x^1+i x^2, \bar{z}=x^1-i x^2$, 
and the inverse of the K\"ahler metric $K^{-1 i j^*}$ 
are introduced. 
We can also consider gauge interactions. 
{}For simplicity we take only the $U(1)$ gauge interaction. 
Then the derivative $\partial A^i / \partial \bar{z}$ in the above 
Eq.(\ref{Be1}) should be replaced by the gauge covariant derivative 
${\cal D}_{\bar{z}} A^i$  
\begin{equation}
2 
{\cal D}_{\bar{z}} A^i
= \Omega_{+}K^{-1 i j^*}\frac{\partial {\cal W}^*}{ \partial A^{*j}}, \quad 
{\cal D}_{\bar{z}} =\frac{1}{2}({\cal D}_1+i {\cal D}_2), \quad 
{\cal D}_\mu A^i = 
\left({\partial \over \partial x^\mu} + i {e_i \over 2} v_\mu
\right)A^i .
\end{equation}
Moreover the same BPS condition (\ref{conseved_charge_1st}) applied to 
vector 
superfield in the Wess-Zumino gauge $V=(v_\mu, \lambda, D)$ gives 
after eliminating the auxiliary field $D$  
\begin{equation}
v_{12}=-D={1 \over 2} \sum_j A^{*j} e_j A^j, \quad v_{03}=0, \quad 
v_{01}=v_{31}, \quad v_{23}=-v_{02}, 
\label{Be1vector}
\end{equation}
where 
$v_{\mu \nu}\equiv 
\partial_{\mu} v_{\nu} - \partial_{\nu} v_{\mu}$ and $e_j$ is the charge 
of the field $A^j$. 
Here we assume for simplicity the minimal 
kinetic term both for 
the chiral superfield $K_{i j^*} = \delta_{i j^*}$ 
and for the vector superfield. 

Similarly the condition of supercharge conservation 
(\ref{conseved_charge_2st}) 
for $H=H_{\rm II}$ applied to chiral superfield  in the Wess-Zumino model 
gives after eliminating the auxiliary field 
\begin{equation}
2 
{\partial A^i \over \partial z} 
=- \Omega_{-} F^i 
= \Omega_{-}K^{-1 i j^*} \frac{\partial {\cal W}^*}{ \partial A^{*j}}, \quad 
\Omega_{-}\equiv i\frac{\langle -i Z_1^*-Z_2^* \rangle}
{|\langle -i Z_1^*-Z_2^* \rangle|} .
\label{Be2}
\end{equation}
If $U(1)$ gauge interaction is present, the derivative 
$\partial A^i/ \partial z$ should be replaced by the covariant derivative 
${\cal D}_z A^i=\frac{1}{2}({\cal D}_1-i {\cal D}_2) A^i$. 
 In this case 
the BPS condition applied to $U(1)$ vector 
superfield in the Wess-Zumino gauge 
 becomes
in the case of minimal kinetic terms
\begin{equation}
v_{12}=D=-{1 \over 2} \sum_j A^{*j} e_j A^j, \quad v_{03}=0, \quad 
v_{01}=-v_{31}, \quad v_{23}=v_{02} . 
\label{Be2vector}
\end{equation}
In sect.\ref{sc:bc-bps}, we shall present a simple way to find the 
correspondence 
between the choice of boundary conditions and the choice of BPS 
equations (\ref{Be1}) and (\ref{Be1vector}) 
or (\ref{Be2}) and (\ref{Be2vector}). 

\subsection{The exact solution of BPS domain wall junction}

\begin{figure}[htbp]
\begin{center}
\leavevmode
\epsfxsize=7cm
\epsfysize=7cm
\centerline{\epsfbox{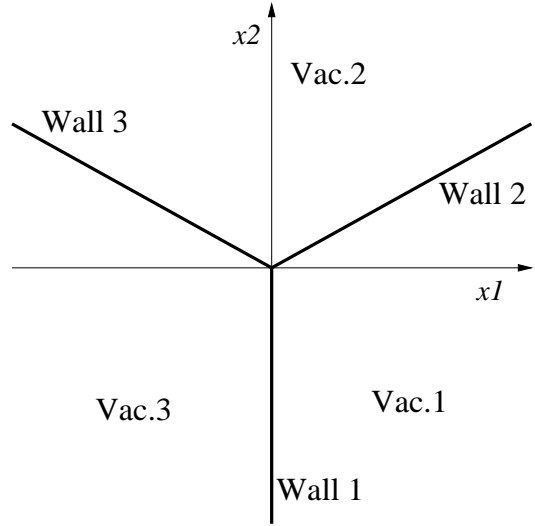}}
\caption{Boundary condition of the model in \cite{HKMN}}
\label{FIG:boundary}
\end{center}
\end{figure}

In a previous article \cite{HKMN}, we have found an 
exact solution of BPS domain wall junction in a model motivated by 
the ${\cal N}=2$ supersymmetric $SU(2)$ gauge theory with one flavor 
 broken to ${\cal N}=1$ by the mass of the adjoint chiral superfield. 
This model has the following chiral superfields with the charge 
assignment for the $U(1)\times U(1)'$ gauge group 
\begin{equation}
\begin{array}{cccccccc}
      & {\cal M} & \tilde{\cal M} & {\cal D} & \tilde{\cal D} 
      & {\cal Q} & \tilde{\cal Q} & T \\
U(1)  &  0       &  0             &   1      &  -1            
      &  1       & -1             &   0 \\
U(1)' &  1       & -1             &   1      &  -1  
      &  0       &  0             & 0  ,
\end{array}
\end{equation}
interacting with a superpotential 
\begin{equation}
{\cal W}=(T-\Lambda) {\cal M} \tilde{{\cal M}}
+ (T+\Lambda) {\cal D} \tilde{{\cal D}}
+ (T-m) {\cal Q} \tilde{{\cal Q}}
- h^2 T,
\end{equation}
where parameters $\Lambda$ and $h$ can be made real positive and a 
 parameter $m$ is complex \cite{HKMN}. 
In this model there are three discrete vacua, 
\begin{eqnarray}
{\rm Vac.1} &\!\!\!:&\!\!\!
T=\Lambda, \quad {\cal M}=\tilde{\cal M} = h, \quad
{\cal Q}=\tilde{\cal Q}={\cal D}=\tilde{\cal D}=0, \quad
{\cal W}_1=-h^2 \Lambda, \nonumber \\
{\rm Vac.2} &\!\!\!:&\!\!\!
T=m, \quad {\cal Q}=\tilde{\cal Q}= h, \quad
{\cal M}=\tilde{\cal M}={\cal D}=\tilde{\cal D}=0, \quad 
{\cal W}_2=-h^2 m, \nonumber \\
{\rm Vac.3} &\!\!\!:&\!\!\!
T=-\Lambda, \quad {\cal D}=\tilde{\cal D}= h, \quad
{\cal Q}=\tilde{\cal Q}={\cal M}=\tilde{\cal M}=0, \quad 
{\cal W}_3=h^2 \Lambda,
\label{vac}
\end{eqnarray}
and when $m=i\sqrt{3}\Lambda$, this model becomes $Z_3$ symmetric. 
Thus three half walls are expected to connect at the junction 
with relative angles of $2\pi/3$.
{}For definiteness, we specify the boundary condition where the 
wall $1$ extends along the negative $x^2$ axis separating the vacuum 
$1$ ($x^1>0$) and $3$ ($x^1<0$) as shown 
in Fig.~\ref{FIG:boundary}.  
If we have only the wall $1$, we obtain the central charge $Z_k$ 
(vanishing $Y_k$) and find 
the two conserved supercharges 
from Eqs.(\ref{conseved_charge_1st}) 
and (\ref{conseved_charge_2st}) 
as 
\begin{equation}
 Q^{(1)} = 
\frac{1}{\sqrt{2}}({\rm e}^{-i{\pi \over 4}} Q_2 
+ {\rm e}^{i{\pi \over 4}} \overline{Q}_{\dot{2}}),
\qquad 
  Q^{(2)} = 
\frac{1}{\sqrt{2}}({\rm e}^{i{\pi \over 4}}Q_1 
+{\rm e}^{-i{\pi \over 4}} \overline{Q}_{\dot{1}}). 
\end{equation}
The other two walls have also two conserved supercharges 
\begin{eqnarray}
\mbox{at wall 2} & Q^{(3)} = \frac{1}{\sqrt{2}} 
                            ({\rm e}^{-i{\pi \over 12}} Q_1 
                          + {\rm e}^{i{\pi \over 12}} \overline{Q}_{\dot{1}}), 
            \,\, 
 \mbox{besides} \,\,\, 
  Q^{(1)} = \frac{1}{\sqrt{2}}
({\rm e}^{-i{\pi \over 4}} Q_2 
+ {\rm e}^{i{\pi \over 4}} \overline{Q}_{\dot{2}}),
             \nonumber \\
\mbox{at wall 3} & 
                   Q^{(4)} = \frac{1}{\sqrt{2}}
                             ({\rm e}^{-i{5\pi \over 12}} Q_1 
                          + {\rm e}^{i{5\pi \over 12}}\overline{Q}_{\dot{1}}), 
           \,\, 
 \mbox{besides} \,\,\,
  Q^{(1)} = \frac{1}{\sqrt{2}}
({\rm e}^{-i{\pi \over 4}} Q_2 
+ {\rm e}^{i{\pi \over 4}} \overline{Q}_{\dot{2}}) .
\label{wallcharge}
\end{eqnarray}
When these three half walls coexist, we can have only one common conserved 
supercharge 
$  Q^{(1)} = ({\rm e}^{-i{\pi \over 4}} Q_2 
+ {\rm e}^{i{\pi \over 4}} \overline{Q}_{\dot{2}})/\sqrt{2}$. 
In fact we find that the domain wall junction configuration conserves 
precisely this single combination of supercharges, 
even though it has also another central charge $Y_k$ contributing. 
Correspondingly we obtain the BPS equations (\ref{Be2}) and (\ref{Be2vector}) 
for $H=H_{{\rm II}}$ with $\Omega_{-}=-1$. 
The BPS equations (\ref{Be2vector}) for the vector superfield can be trivially 
satisfied  by $v_{\mu}=0$ and $D=0$. 
The BPS equations (\ref{Be2}) for chiral superfields become in this case 
\begin{equation}
2 \frac{ \partial A^i }{ \partial z}
=  -\frac{\partial {\cal W}^*}{\partial A^{*i}} ,
\label{BPSeqs}
\end{equation}
assuming the minimal kinetic term. 
The solution for these BPS equations is given by \cite{HKMN},
\begin{eqnarray}
{\cal M}(z, \bar{z})
&\!\!\! = &\!\!\! 
\tilde{\cal M}(z, \bar{z})=
\frac{\sqrt2 \Lambda s}{s+t+u}, \nonumber \\
{\cal D}(z, \bar{z})
&\!\!\! = &\!\!\! 
\tilde{\cal D}(z, \bar{z})=
\frac{\sqrt2 \Lambda t}{s+t+u}, \nonumber \\
{\cal Q}(z, \bar{z})
&\!\!\! = &\!\!\! 
\tilde{\cal Q}(z, \bar{z})=
\frac{\sqrt2 \Lambda u}{s+t+u}, \nonumber \\
T(z, \bar{z})
&\!\!\! = &\!\!\! 
\frac{2 \Lambda}{\sqrt{3}}
\frac{
e^{-i \frac{1}{6} \pi} s
+ e^{-i \frac{5}{6} \pi} t
+ e^{i \frac{1}{2} \pi} u
}{s+t+u}+\frac{i}{\sqrt3}\Lambda, 
\label{exact solution}
\end{eqnarray}
\begin{eqnarray}
s  = \exp \left( \frac{2 \Lambda}{\sqrt{3}}
{\rm Re} \left( e^{i \frac{1}{6} \pi} z \right) \right),
\quad 
t  = \exp \left( \frac{2 \Lambda}{\sqrt{3}}
{\rm Re} \left( e^{i \frac{5}{6} \pi} z \right) \right),
\quad 
u  =  \exp \left( \frac{2 \Lambda}{\sqrt{3}}
{\rm Re} \left( e^{-i \frac{1}{2} \pi} z \right) \right).
\label{stu}
\end{eqnarray}

This model is motivated by the softly broken ${\cal N}=2$ $SU(2)$ 
gauge theory with one flavor. 
However, we can simplify the model without spoiling the solvability 
to obtain a Wess-Zumino model consisting of purely chiral superfields 
by the following procedure. 
The vector superfields actually serve to constrain chiral superfields 
to have the identical magnitude pairwise through $D=0$ 
to satisfy the BPS equation (\ref{Be2vector}) for vector superfields: 
$|\tilde{\cal M}|=|{\cal M}|, |\tilde{\cal D}|=|{\cal D}|, 
|\tilde{\cal Q}|=|{\cal Q}|$. 
Therefore we can eliminate the vector superfields and reduce the number of 
chiral superfields by identifying pairwise 
$\tilde{\cal M}={\cal M}, \tilde{\cal D}={\cal D}, 
\tilde{\cal Q}={\cal Q}$. 
Correspondingly we should take the superpotential as 
\begin{equation}
{\cal W}={1 \over 2}(T-\Lambda) {\cal M}^2
+ {1 \over 2}(T+\Lambda) {\cal D}^2
+ {1 \over 2}(T-i\sqrt{3}\Lambda) {\cal Q}^2
- {h^2 \over 2} T. 
\end{equation}
This Wess-Zumino model has the same solution as ours 
by changing $h^2 \rightarrow h^2/2, \Lambda \rightarrow \sqrt3 \Lambda/2$. 
A similar observation has also been made in ref.\cite{ShifmanVeldhuis}. 

\subsection{Unitary representations of $(1, 0)$ supersymmetry algebra}
\label{sc:unitaryrep}
Let us examine states on the background of a domain wall junction from 
the point of view of surviving symmetry. 
In the case of the BPS states satisfying the BPS equation (\ref{Be2}) 
corresponding to $H=H_{{\rm II}}$, we 
have only one surviving supersymmetry charge $Q^{(1)}$, 
two translation generators $H, P^3$, and one Lorentz generator $J^{0 3}$, 
out of the ${\cal N}=1$ four dimensional super Poincar\'{e} generators. 
Since we are interested in excitation modes on the background of the 
domain wall junction, we define the hamiltonian $H'=H-\langle H \rangle$ 
measured from the energy $\langle H \rangle$ of the background configuration. 
By projecting from the supersymmetry algebra (\ref{qqantcom}), 
(\ref{qqantcomy}) with central charges in four dimensions, 
we immediately find 
\begin{equation}
\left(Q^{(1)}\right)^2=H' -P^3 . 
\end{equation}
We also obtain the Poincar\'{e} algebra in $1+ 1$ dimensions 
\begin{equation}
[J^{03}, Q^{(1)}]= {i \over 2} Q^{(1)}, \quad 
[J^{03}, H'-P^3]= i (H'-P^3), \quad 
[J^{03}, H'+P^3]= -i (H'+P^3) . 
\end{equation}
Other commutation relations are trivial 
\begin{equation}
[H'-P^3, H'+P^3]=[H'-P^3, Q^{(1)}]= 
[H'+P^3, Q^{(1)}]=0 . 
\end{equation}
This is precisely the $(1, 0)$ supersymmetry algebra on the domain wall 
junction as anticipated \cite{GibbonsTownsend}. 

To obtain unitary representations, we can diagonalize $H'$ and $P^3$ 
\begin{equation}
H'\vert E, p^3 \rangle =E \vert E, p^3 \rangle, \qquad 
P^3 \vert E, p^3 \rangle =p^3 \vert E, p^3 \rangle, \qquad E \ge |p^3|,
\end{equation}
and combine them by means of $Q^{(1)}$. 
If $E-p^3 > 0$, we can construct bosonic state from fermionic state and vice 
versa by operating $Q^{(1)}$ on the state. 
\begin{equation}
\vert B \rangle ={1 \over \sqrt{E-p^3}}Q^{(1)} \vert F \rangle, \qquad 
\vert F \rangle ={1 \over \sqrt{E-p^3}}Q^{(1)} \vert B \rangle.
\end{equation}
Therefore we obtain a doublet representation
 $( \vert B \rangle, \vert F \rangle )$. 
If $E-p^3 = 0$, operating by 
 $Q^{(1)}$ on the state gives an unphysical zero norm state 
\begin{equation}
\left|Q^{(1)}\vert E, p^3 \rangle\right|^2 
=\langle E, p^3 \vert \left(Q^{(1)}\right)^2 \vert E, p^3 \rangle 
=\langle E, p^3 \vert H'-P^3 \vert E, p^3 \rangle 
=E-p^3=0.
\end{equation}
Then the massless right-moving state $\vert E, p^3=E \rangle$ 
is  a singlet representation. 
This singlet state can either be boson or fermion. 
Thus we find that there are only two types of representations of the 
 $(1, 0)$ supersymmetry algebra, doublet and singlet. 
We also find that massive modes should appear in pairs of boson and fermion, 
whereas the massless right-moving mode can appear singly without 
accompanying a state with opposite statistics. 
This provides an interesting possibility of a chiral structure for 
fermions. 

If another BPS equation (\ref{Be1}) corresponding to $H=H_{\rm I}$ is 
satisfied instead of Eq. (\ref{Be2}), we have $(0, 1)$ 
supersymmetry and the left-moving massless states can appear as 
singlets.

\section{Nambu-Goldstone and other modes on the junction}

\subsection{Mode equation on the junction}

Since the vector superfields have no nontrivial field configurations, 
Nambu-Goldstone modes have no component of vector superfield. 
Moreover we can replace our model, if we wish, by another 
model with purely chiral superfields without spoiling the 
essential features including the solvability. 
Consequently we shall neglect 
vector superfields and consider the general Wess-Zumino model 
in Eq.(\ref{generalWZmodel}) in the following. 
{}For simplicity we assume the minimal kinetic term 
here $K_{ij^*}=\delta_{ij^*}$ .

Let us consider 
quantum fluctuations $A'^i, \psi^i$ 
around a classical solution $A^i_{\rm cl}$ which satisfies the BPS equations 
 (\ref{Be1}) and (\ref{Be1vector}) for $H=H_{\rm I}$ 
or (\ref{Be2}) and (\ref{Be2vector}) for $H=H_{\rm II}$. 
\begin{equation}
A^i=A^i_{\rm cl} + A'^i .
\end{equation}
We retain the part of the Lagrangian quadratic in fluctuations 
and eliminate the auxiliary fields $F^i$ to obtain the linearized 
equation for the scalar fluctuations 
\begin{equation}
-\partial_\mu \partial^\mu A'^{*i}
+{\partial^2 {\cal W} \over \partial A^{i}_{\rm cl}\partial A^{k}_{\rm cl}}
{\partial^2 {\cal W}^* \over \partial A^{*k}_{\rm cl}\partial A^{*j}_{\rm cl}}
 A'^{*j}
+{\partial^3 {\cal W} \over 
\partial A^{i}_{\rm cl}\partial A^{k}_{\rm cl}\partial A^{j}_{\rm cl}}
{\partial {\cal W}^* \over \partial A^{*k}_{\rm cl}}
 A'^{j}
 = 0 .
\end{equation}
In order to separate variables 
in $x^0, x^3$ and $x^1, x^2$ we have 
to define mode equations on the background which has a nontrivial 
dependence in two dimensions, $x^1, x^2$. 
The bosonic modes $ A'^{i}_n(x^1, x^2)$ can easily be defined in terms of 
a differential operator ${\cal O}_B$ in $x^1, x^2$ space 
\begin{equation}
{\cal O}_B{}^i{}_j \equiv 
\left[\matrix{
-\left(\partial_1^2 + \partial_2^2\right) \delta^i{}_j 
+{\partial^2 {\cal W} \over \partial A^{i}_{\rm cl}\partial A^{k}_{\rm cl}}
{\partial^2 {\cal W}^* \over \partial A^{*k}_{\rm cl}\partial A^{*j}_{\rm cl}}
& 
{\partial^3 {\cal W} \over 
\partial A^{i}_{\rm cl}\partial A^{k}_{\rm cl}\partial A^{j}_{\rm cl}}
{\partial {\cal W}^* \over \partial A^{*k}_{\rm cl}}
\cr
{\partial^3 {\cal W}^* \over 
\partial A^{*i}_{\rm cl}\partial A^{*k}_{\rm cl}\partial A^{*j}_{\rm cl}}
{\partial {\cal W} \over \partial A^{k}_{\rm cl}}
& 
-\left(\partial_1^2 + \partial_2^2\right) \delta^{i}{}_j 
+{\partial^2 {\cal W}^* \over \partial A^{*i}_{\rm cl}\partial A^{*k}_{\rm cl}}
{\partial^2 {\cal W} \over \partial A^{k}_{\rm cl}\partial A^{j}_{\rm cl}}
\cr}\right]
\end{equation}
\begin{equation}
{\cal O}_B{}^i{}_j 
\left[\matrix{
A'^{*j}_n \cr 
 A'^{j}_n
\cr}\right]
 = M_n^2  
\left[\matrix{
A'^{*i}_n \cr 
 A'^{i}_n
\cr}\right] ,
\end{equation}
where the eigenvalue $M_n^2$ has to be real from Majorana condition. 
The quantum fluctuation for scalar can be expanded in terms of these 
mode functions to obtain a real scalar field equation with the mass $M_n$ 
for the coefficient bosonic field $a_n(x^0, x^3)$ 
\begin{equation}
 A'^{i}(x^0, x^1, x^2, x^3)=\sum_n a_n(x^0, x^3) A'^i_n(x^1, x^2)
\end{equation}
\begin{equation}
 \left(\partial_0^2 -\partial_3^2 + M_n^2 \right) a_n(x^0, x^3) =0 .
\end{equation}

Similarly the linearized equation for fermions is given by 
\begin{equation}
-i \bar \sigma^\mu \partial_\mu \psi^i - 
{\partial^2 {\cal W}^* \over \partial A^{*i}_{\rm cl}\partial A^{*j}_{\rm cl}}
\bar \psi^j = 0
\label{eq:lin-fermion1}
\end{equation}
\begin{equation}
-i \sigma^\mu \partial_\mu \bar \psi^i - 
{\partial^2 {\cal W} \over \partial A^{i}_{\rm cl}\partial A^{j}_{\rm cl}}
 \psi^j = 0 .
\label{eq:lin-fermion2}
\end{equation}
To separate variables for fermion equations, it is more convenient to use 
a gamma matrix representation where direct product structure of $2 \times 2$ 
matrices for $(x^0,x^3)$ and $(x^1, x^2)$ space is manifest. 
We shall describe one such representation in appendix B. 
Transforming from such a representation to the Weyl representation 
which we are using, we can define the fermionic 
modes $\psi^i_{n \alpha}, \bar \psi^{i \dot \beta}_n$ 
combining components of left-handed and right-handed spinors 
by means of the following operators 
\begin{equation}
{\cal O}_1{}^i{}_j \equiv 
\left[\matrix{
- {\partial^2 {\cal W}^* \over \partial A^{*i}_{\rm cl}\partial A^{*j}_{\rm cl}}
&
-i\left(-\partial_1+i\partial_2\right) \delta^i_j \cr
-i\left(\partial_1+i\partial_2\right) \delta^i_j 
&
- {\partial^2 {\cal W} \over \partial A^{i}_{\rm cl}\partial A^{j}_{\rm cl}}
}\right]
\label{eq:fermion-mode-weyl1}
\end{equation}
\begin{equation}
{\cal O}_2{}^i{}_j \equiv 
\left[\matrix{
- {\partial^2 {\cal W} \over \partial A^{i}_{\rm cl}\partial A^{j}_{\rm cl}}
&
-i\left(\partial_1-i\partial_2\right) \delta^i_j \cr
-i\left(-\partial_1-i\partial_2\right) \delta^i_j 
&
- {\partial^2 {\cal W}^* \over \partial A^{*i}_{\rm cl} 
\partial A^{*j}_{\rm cl}}
}\right]
\label{eq:fermion-mode-weyl2}
\end{equation}
\begin{equation}
{\cal O}_1{}^i{}_j 
\left[\matrix{\bar \psi^{j \dot 1}_n \cr \psi^j_{n 2} \cr}\right]
 = -i m_n^{(1)} 
\left[\matrix{ \psi^i_{n 1} \cr \bar \psi^{i \dot 2}_n \cr}\right]
\label{eq:fermion-mode-weyl3}
\end{equation}
\begin{equation}
{\cal O}_2{}^i{}_j 
\left[\matrix{\psi^j_{n 1} \cr \bar \psi^{j \dot 2}_n \cr}\right]
 = i m_n^{(2)} 
\left[\matrix{\bar \psi^{i \dot 1}_n \cr \psi^i_{n 2} \cr}\right] ,
\label{eq:fermion-mode-weyl4}
\end{equation}
where the mass eigenvalues $m_n^{(1)}, m_n^{(2)}$ are real. 
Please note a peculiar combination of left- and right-handed 
spinor components to define eigenfunctions. 
We can expand $\psi^i$ in terms of these mode functions 
\begin{equation}
 \psi^i_\alpha (x^0, x^1, x^2, x^3)=\sum_n 
\left(\matrix{b_n(x^0, x^3) \psi^i_{n1}(x^1, x^2)
\cr 
c_n(x^0, x^3) \psi^i_{n2}(x^1, x^2)\cr}\right)
\end{equation}
Since $\psi(x^0,x^1,x^2,x^3)$ is a Majorana spinor, the coefficient 
fermionic fields $b_n, c_n$ are real. 
The linearized equations (\ref{eq:lin-fermion1}) (\ref{eq:lin-fermion2}) 
for the fermion gives a Dirac equation 
in $1+1$ dimensions for the coefficient fermionic fields $(c_n, i b_n)$ 
with two mass parameters $m_n^{(1)}, m_n^{(2)}$ 
\begin{equation}
 \left[-i\left(\rho_1\partial_0 + i \rho_2\partial_3\right) 
-m_n^{(1)}{1 + \rho_3 \over 2} 
-m_n^{(2)}{1 - \rho_3 \over 2} \right]
\left[\matrix{ c_n(x^0, x^3) \cr i b_n(x^0, x^3)\cr}\right] =0 ,
\label{eq:dirac1+1}
\end{equation}
where we use Pauli matrices $\rho_a,  \ a=1,2,3$ to construct the 
$2 \times 2$ gamma matrices $\rho_1, i\rho_2$ in $1+1$ dimensions. 
Since we have a Majorana spinor in $1+1$ dimensions which does not allow 
 chiral rotations, we have 
two distinct real mass parameters $m_n^{(1)}, m_n^{(2)}$. 

To relate the mass eigenvalues of fermions and bosons, 
let us multiply two differential operators for fermions 
${\cal O}_2 $ to ${\cal O}_1$. 
In this ordering, we can use the BPS equation (\ref{Be2}) corresponding 
to $H=H_{{\rm II}}$ to find the differential operator for bosons ${\cal O}_B$ 
\begin{equation}
{\cal O}_2^i{}_k 
{\cal O}_1^k{}_j 
 = \left[\matrix{{\rm e}^{i{\pi \over 4}}\Omega_-^{1\over 2} & 0\cr 
                 0 & {\rm e}^{-i{\pi \over 4}}\Omega_-^{-{1\over 2}}\cr}\right]
{\cal O}_B^i{}_j 
  \left[\matrix{{\rm e}^{-i{\pi \over 4}}\Omega_-^{-{1\over 2}} & 0\cr 
                 0 & {\rm e}^{i{\pi \over 4}}\Omega_-^{1\over
                   2}\cr}\right] .
\end{equation}
Therefore the BPS equation (\ref{Be2}) corresponding 
to $H=H_{{\rm II}}$ guarantees that the existence of a solution 
$\bar \psi_n^{i\dot 1}, \psi_{n 2}^i$ 
of fermionic mode equations implies the existence of a solution of 
bosonic mode equations with the mass squared $M_n^2=m_n^{(1)}m_n^{(2)}$ 
\begin{equation}
A_n'^{*i}
={\rm e}^{-i{\pi \over 4}}\Omega_-^{-{1\over 2}}\bar \psi_n^{i\dot 1} ,
\qquad 
A_n'^{i}={\rm e}^{i{\pi \over 4}}\Omega_-^{{1\over 2}} \psi_{n2}^{i} . 
\end{equation}

If another 
BPS equation (\ref{Be1}) corresponding to $H=H_{\rm I}$ is valid, 
operator multiplication with different ordering gives the 
same bosonic operator whose rows and columns are interchanged 
\begin{equation}
{\cal O}_1^i{}_k 
{\cal O}_2^k{}_j 
 = \left[\matrix{0 & {\rm e}^{i{\pi \over 4}}\Omega_+^{-{1\over 2}} \cr 
                 -{\rm e}^{-i{\pi \over 4}}\Omega_+^{1\over 2} & 0 \cr}\right]
{\cal O}_B^i{}_j 
   \left[\matrix{0 & -{\rm e}^{i{\pi \over 4}}\Omega_+^{-{1\over 2}} \cr 
                 {\rm e}^{-i{\pi \over 4}}\Omega_+^{1\over 2} & 0
                 \cr}\right] .
\end{equation}
Therefore the BPS equation (\ref{Be1}) corresponding 
to $H=H_{\rm I}$ guarantees that the existence of a solution 
$\bar \psi_n^{i\dot 2}, \psi_{n1}^{i}$ 
of fermionic mode equations implies the existence of a solution of 
bosonic mode equations with the mass squared $M_n^2=m_n^{(1)}m_n^{(2)}$  
\begin{equation}
A_n'^{*i}
=-{\rm e}^{i{\pi \over 4}}\Omega_+^{-{1\over 2}}\bar \psi_n^{i\dot 2} , 
\qquad 
A_n'^{i}={\rm e}^{-i{\pi \over 4}}\Omega_+^{{1\over 2}} \psi_{n1}^{i} . 
\end{equation}

Therefore we find that all massive states come in pairs of boson and 
fermion with the same mass squared $M_n^2=m_n^{(1)}m_n^{(2)}$ 
in accordance with the result of the unitary representation of 
the $(1, 0)$ supersymmetry 
algebra. 


\subsection{Nambu-Goldstone modes}

Since we are usually most interested in a low energy effective field theory, 
we wish to study massless modes here. 
If global continuous symmetries are broken spontaneously, 
there occur associated massless modes which are called the Nambu-Goldstone 
modes. 
To find the wave functions of the Nambu-Goldstone modes, we perform 
the associated global transformations and evaluate 
the transformed configuration by substituting the classical field. 
For supersymmetry we obtain nontrivial wave function by substituting 
the classical field $A^i_{\rm cl}(x^1,x^2)$ and $F^i_{\rm cl}(x^1,x^2)$ 
to the transformation of fermions by a Grassmann parameter $\xi$, since 
classical field configuration of fermion vanishes $\psi^i_{\rm cl}=0$ 
\begin{equation}
\delta_{\xi} \psi^i = i\sqrt{2}\sigma^\mu\bar{\xi}\partial_\mu A^i_{\rm cl} 
+ \sqrt{2} \xi F^i_{\rm cl} .
\end{equation}
If the BPS equation (\ref{Be2}) for the junction background is valid, 
we obtain 
\begin{equation}
\delta_{\xi} \psi^i =  \sqrt{2}\left[ 
(i\sigma^1\bar{\xi}-\Omega^*_{-}\xi)\partial_1 A^i_{\rm cl}
+(i\sigma^2\bar{\xi}+i\Omega^*_{-}\xi)\partial_2 A^i_{\rm cl}
\right]. 
\end{equation}
We see that there is one conserved direction in the Grassmann parameter: 
\begin{equation}
i\sigma^1\bar{\xi}\, =\, \Omega^*_{-}\xi \,\,\, \mbox{and} \,\,\, 
\sigma^2\bar{\xi}\, =\, -\Omega^*_{-}\xi .
\end{equation}
The other three real Grassmann parameters $\xi$ correspond to 
broken supercharges. 
{}For our exact solution, for instance, we find it convenient to choose 
the three broken supercharges as the following real supercharges 
\begin{equation}
  Q_{\rm I}= \frac{1}{\sqrt{2}}
              (e^{i\pi/4}Q_2 + e^{-i\pi/4}\bar{Q}_{\dot{2}}), \,\, 
  Q_{\rm II}= \frac{1}{\sqrt{2}}
              (e^{-i\pi/4}Q_1 + e^{i\pi/4}\bar{Q}_{\dot{1}}), \,\,
  Q_{\rm III}= \frac{1}{\sqrt{2}}
              (e^{i\pi/4}Q_1 + e^{-i\pi/4}\bar{Q}_{\dot{1}}).
\label{eq:brokencharges}
\end{equation} 
Then the corresponding massless mode functions are given by 
\begin{eqnarray}
  \psi_0^{{\rm (I)}i}(x^1,x^2) 
    &=&
    \left( 
    \begin{array}{c}       
   4\partial_z A_{\rm cl}^i(x^1,x^2) e^{-i\pi/4} \\
       0 
    \end{array}
    \right),    
  \label{NG1} \\
  \psi_0^{{\rm (II)}i}(x^1,x^2)
    &=&  
    \left( 
    \begin{array}{c}
       0 \\
   2\partial_1 A_{\rm cl}^i(x^1,x^2)e^{i\pi/4} 
    \end{array}
    \right),
  \label{NG2} \\
  \psi_0^{{\rm (III)}i}(x^1,x^2)
    &=&  
    \left( 
    \begin{array}{c}
       0 \\ 
       2\partial_2 A_{\rm cl}^i(x^1,x^2) e^{i\pi/4} 
    \end{array}
    \right).
  \label{NG3} 
\end{eqnarray}
Since the transformation parameter should correspond to the Nambu-Goldstone 
field with zero momentum and energy, the three transformation parameters $\xi$ 
should be promoted to three real fermionic fields in $x^0, x^3$ space, 
 $b_0^{({\rm I})}(x^0, x^3), c_0^{({\rm II})}(x^0, x^3)$, and 
$c_0^{({\rm III})}(x^0, x^3)$, 
to obtain the Nambu-Goldstone component of the mode expansion 
\begin{eqnarray}
  \psi^{i}(x^0, x^1,x^2, x^3) 
    &=& b_0^{({\rm I})}(x^0, x^3)   \psi_0^{{\rm ({\rm I})}i}(x^1,x^2) 
+ c_0^{({\rm II})}(x^0, x^3)   \psi_0^{{\rm ({\rm II})}i}(x^1,x^2) 
\nonumber \\
    && + c_0^{({\rm III})}(x^0, x^3)   \psi_0^{{\rm ({\rm III})}i}(x^1,x^2) 
+ \sum_{n > 0} 
\left(\matrix{b_n(x^0, x^3) \psi^i_{n1}(x^1, x^2)
\cr 
c_n(x^0, x^3) \psi^i_{n2}(x^1, x^2)\cr}\right).
\end{eqnarray}
We have explicitly displayed three massless Nambu-Goldstone fermion components 
 distinguishing from the massive ones ($n>0$). 
The Dirac equation for the coefficient fermionic fields (\ref{eq:dirac1+1}) 
shows that  $b_0^{({\rm I})}(x^0-x^3)$ is a right-moving massless mode, 
and $c_0^{({\rm II})}(x^0+x^3)$, and 
$c_0^{({\rm III})}(x^0+x^3)$ are left-moving modes. 

\begin{figure}[htbp]
\begin{flushleft}
\vspace{3cm}
\leavevmode
\begin{eqnarray*}
\begin{array}{cc}
  \epsfxsize=5.5cm
  \epsfysize=4.5cm
  \hspace{3cm}
  \epsfbox{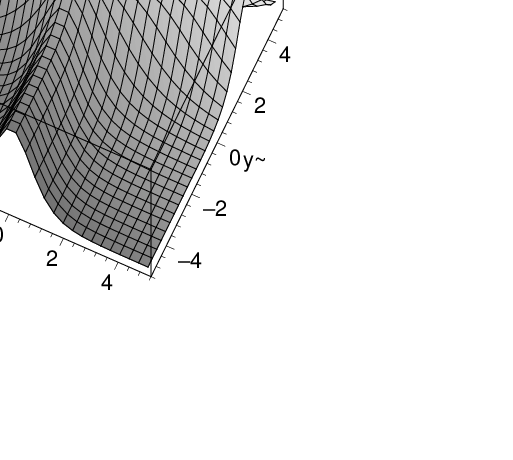} &
  \epsfxsize=5.5cm
  \epsfysize=4.5cm
  \hspace{3cm}
  \epsfbox{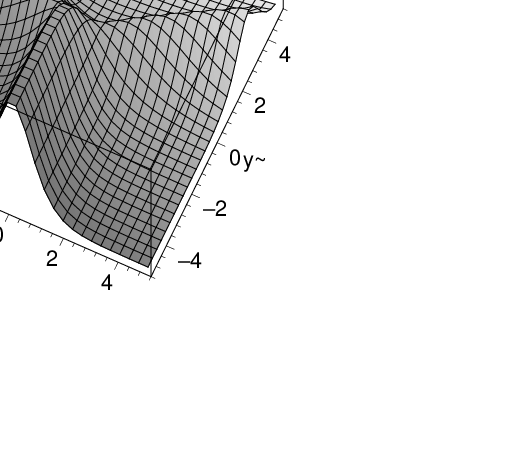} \\
  \mbox{The wave function $\left|\psi^{{\rm (I)}T}_0\right|$ } \quad  & 
  \mbox{The wave function $\left|\psi^{{\rm (II)}T}_0\right|$} \quad \\
\end{array} 
\end{eqnarray*} 
\vspace{2cm} 
\begin{eqnarray*}
\begin{array}{cc}
  \epsfxsize=5.5cm
  \epsfysize=4.5cm
  \hspace{3cm}
  \epsfbox{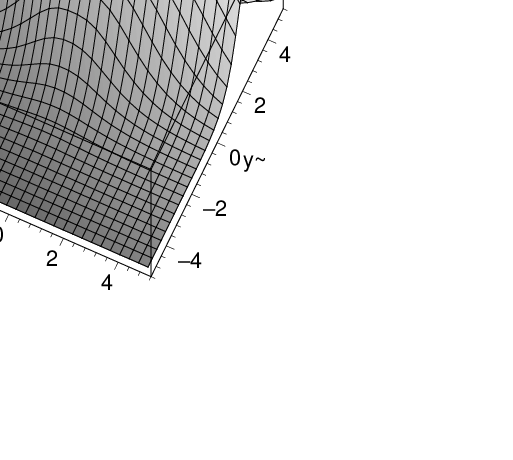} &
  \epsfxsize=5.5cm
  \epsfysize=4.5cm
  \hspace{8.5cm} \\
  \mbox{The wave function $\left|\psi^{{\rm (III)}T}_0\right|$} \quad & 
\end{array}
\end{eqnarray*}
\caption{The bird's eye view of the absolute value of the $i=T$ component 
of the wave functions of the Nambu-Goldstone fermions 
on the junction in the $(x^1,x^2)$ space
}
\label{FIG:NGf}
\end{flushleft}
\end{figure}

We plot the absolute values of $|\psi_0^{(a) i=T}|$ of the $i=T$ component 
of the wave function of the Nambu-Goldstone fermions 
$a={\rm I}, {\rm II}, {\rm III}$ 
in Fig.~\ref{FIG:NGf}. 
We can see that Nambu-Goldstone fermions have wave 
functions which extend to infinity along three walls. 
They become identical to fermion zero modes on at least two of the walls 
asymptotically 
and hence they are not localized around the center of the junction. 
We can construct a linear combination of the Nambu-Goldstone fermions 
to have no support along one out of the three walls. 
However, no linear combination of these Nambu-Goldstone fermions 
can be formed which does not have support extended along any of the 
wall. 
Therefore these wave functions are not localized and are not normalizable. 
This fact means that 
the low energy dynamics of BPS junction cannot be described by a $1+1$
dimensional effective field theory with a discrete particle spectrum. 

Similarly the Nambu-Goldstone bosons corresponding to the broken translation 
$P^a, a=1, 2$ are given by  
\begin{equation}
A_0^{(a)}(x^1, x^2) = \partial_a A_{\rm cl}^i(x^1, x^2), \qquad a=1, 2. 
\end{equation}
These two bosonic massless modes consist of two left-moving modes and 
two right-moving modes. 
On the other hand, we have seen already that there are two left-moving 
massless Nambu-Goldstone fermions and one right-moving massless 
Nambu-Goldstone fermion. 
These two left-moving Nambu-Goldstone bosons and fermions form 
two doublets of the $(1,0)$ supersymmetry algebra. 
The right-moving modes are asymmetric in bosons and fermions: 
two Nambu-Goldstone bosons and a single Nambu-Goldstone fermion. 
These three states are all singlets of the $(1,0)$ supersymmetry algebra 
in accordance with our analysis in sect.\ref{sc:unitaryrep}. 
Therefore we obtained a chiral structure of Nambu-Goldstone fermions 
on the junction background configuration. 

\subsection{
Non-normalizability of the Nambu-Goldstone 
fermions 
}

We would like to argue that our observation is a generic feature 
of the Nambu-Goldstone fermions on the domain wall junction in a flat 
space in the bulk: Nambu-Goldstone fermions 
are not localized 
at the junction and hence are not normalizable, 
if they are associated with the supersymmetry 
breaking due to the coexistence of nonparallel domain walls. 
The following observation is behind this assertion. 
A single domain wall breaks only a half of supercharges. 
Nonparallel wall also breaks half of supercharges, some of which 
may be linear combinations of the supercharges already broken by the 
first wall. 
If the junction configuration is a $1/4$ BPS state, 
linearly independent ones among these two sets of broken supercharges 
of nonparallel walls become ${3 \over 4}$ of the original supercharges. 

To see in more detail, let us first note that the junction configuration 
reduces asymptotically to a wall if one goes along the wall, say the 
wall $1$. 
On this first wall, a half of the original supersymmetry 
($Q^{(1)}, \cdots, Q^{(N)}$) is broken. 
Denoting the number of original supercharges to be $N$, we call 
these broken supercharges as $Q^{(1)}, \cdots, Q^{(N/2)}$. 
Consequently we have Nambu-Goldstone fermions localized around 
the core of the wall and is constant along the wall. 
In the junction configuration, we have other walls which are not parallel 
to the first wall. 
Asymptotically far away along one of such walls, say wall $2$, 
another half of the 
supersymmetry  $Q'^{(1)}, \cdots, Q'^{(N/2)}$ is broken. 
If the junction is a $1/4$ BPS state, a half of these, 
say $Q'^{(1)}, \cdots, Q'^{(N/4)}$, is a linear combination of  
$Q^{(1)}, \cdots, Q^{(N/2)}$ broken already on the wall $1$. 
The other half,  $Q'^{({N \over 4}+1)}, \cdots, Q'^{({N\over 2})}$ 
are unbroken on the wall $1$. 
Altogether a quarter of the original supercharges remain unbroken. 
Consequently the Nambu-Goldstone fermions corresponding to 
 $Q'^{(1)}, \cdots, Q'^{(N/4)}$ have a wave function which extends to infinity 
and approaches a constant profile along both the walls $1$ and $2$. 
Those modes corresponding to 
 $Q'^{({N \over 4}+1)}, \cdots, Q'^{({N\over 2})}$ 
have support only along the wall $2$, and those corresponding to 
the linear combinations of $Q^{(1)}, \cdots, Q^{(N/2)}$  
orthogonal to $Q'^{(1)}, \cdots, Q'^{(N/4)}$ have support only along the wall 
$1$. 
Thus we find that any linear combinations of the Nambu-Goldstone fermions 
have to be infinitely extended along at least one of the walls which form 
the junction configuration. 
Therefore the Nambu-Goldstone fermions associated with the coexistence 
of nonparallel domain walls are not localized at the junction and 
are not normalizable. 

In our exact solution, domain wall junction configuration reduces 
asymptotically to the wall $1$ at $x^2 \to -\infty$ with fixed $x^1$. 
On the wall, only two supercharges in Eq.(\ref{eq:brokencharges}) are broken 
\begin{equation}
 Q_{\rm I} = \frac{1}{\sqrt{2}}
              (e^{i\pi/4}Q_2 + e^{-i\pi/4}\overline{Q}_{\dot{2}}), 
\qquad 
 Q_{\rm II} = \frac{1}{\sqrt{2}}
              (e^{-i\pi/4}Q_1 + e^{i\pi/4}\overline{Q}_{\dot{1}}) ,
\end{equation}
and there are two corresponding Nambu-Goldstone fermions which become 
domain wall zero modes asymptotically 
\begin{eqnarray}
    \psi_0^{({\rm I})i}(x^1, x^2)&\!\!=&\!\!
    \left( 
    \begin{array}{c}
      4\partial_z A_{\rm cl}^i(x^1,x^2) e^{-i\pi/4} \\
       0 
    \end{array}
    \right) 
    \rightarrow 
    \left( 
    \begin{array}{c}
      2\partial_1 A_{\rm cl}^{i{\rm wall}}(x^1) e^{-i\pi/4} \\
       0 
    \end{array}
    \right), \nonumber \\
    \psi_0^{({\rm II})i}(x^1,x^2)&\!\!=&\!\!
    \left( 
    \begin{array}{c}
       0 \\  
       2\partial_1 A_{\rm cl}^i(x^1,x^2) e^{i\pi/4}
    \end{array}
    \right)
    \rightarrow 
    \left( 
    \begin{array}{c}
       0 \\  
       2\partial_1 A_{\rm cl}^{i{\rm wall}}(x^1) e^{i\pi/4}
    \end{array}
    \right).
\label{NG1-1}
\end{eqnarray}
These wave functions are localized on the core of the wall 1 
in the $x^1$ direction and are constant along the wall. 
Along the other walls we find two broken supercharges one of which 
is identical to one of the broken supercharges, $Q_{\rm I}$. 
The other broken supercharge is $Q'_{\rm II}$ on the wall 2 
and  $Q''_{\rm II}$ on the wall $3$. 
There are only two independent supercharges among 
 $Q_{\rm II}$,  $Q'_{\rm II}$, and  $Q''_{\rm II}$. 
Together with  $Q_{\rm I}$ we obtain three independent broken supercharges. 
We can construct a linear combination of the Nambu-Goldstone fermions to 
have no support along one out of the three walls. 
However, any linear combination has nonvanishing wave function which becomes 
fermion zero mode on at least one of the wall asymptotically. 
Therefore the associated Nambu-Goldstone fermions have support 
which is infinitely extended at least along two of the walls. 

If a single wall is present, we can explicitly construct 
a plane wave solution propagating along the wall, 
which may be called a spin wave and is among massive modes on the wall 
background. 
Even if there are several walls forming a junction configuration, 
we can consider excitation modes which reduce to the spin wave modes 
along each wall. 
They should be a massive mode on the domain wall junction background. 
The Nambu-Goldstone mode on the domain wall junction is 
the zero wave number limit of such a spin wave mode. 
This physical consideration suggests that the massless Nambu-Goldstone 
fermion is precisely the vanishing wave number (along the wall) 
limit of the massive spin wave mode. 

Let us note that our argument does not apply to 
models with the bulk cosmological constant.
In such models, massless graviton is localized on the background
of intersection of walls \cite{ADDK}.
In that case, massless mode is a distinct mode different from the massless 
limit of the massive continuum, although the massless mode is buried 
at the tip of the continuum of massive modes. 
The normalizability of the massless graviton is guaranteed by 
the Anti de Sitter geometry away from the junction or intersection 
including the direction along the wall. 

\section{Boundary conditions and  central charges}

\label{sc:bc-bps}

For a $1/4$ BPS state, there are 
 two sets of BPS equations,
Eqs.(\ref{Be1})--(\ref{Be1vector}) and (\ref{Be2})--(\ref{Be2vector}), 
corresponding to the two kinds of BPS domain wall junctions.  
In this section we make explicit the relation between the boundary 
conditions and the choice of these BPS equations. 

\begin{figure}[htbp]
\begin{center}
\leavevmode
\begin{eqnarray*}
\begin{array}{cc}
\epsfxsize=6cm
\hspace{1cm}
\epsfbox{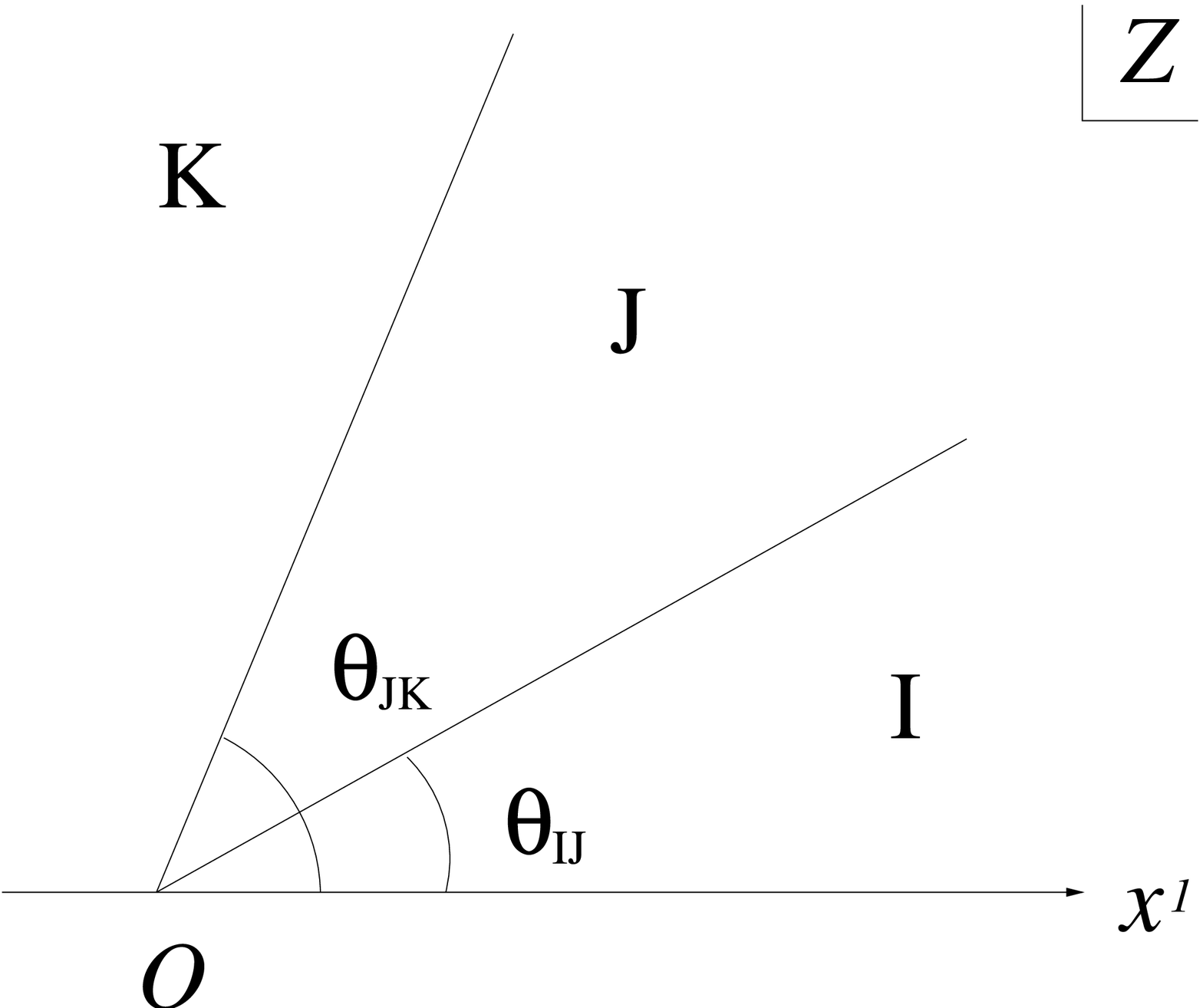} & 
\epsfxsize=6cm
\hspace{1cm}
\epsfbox{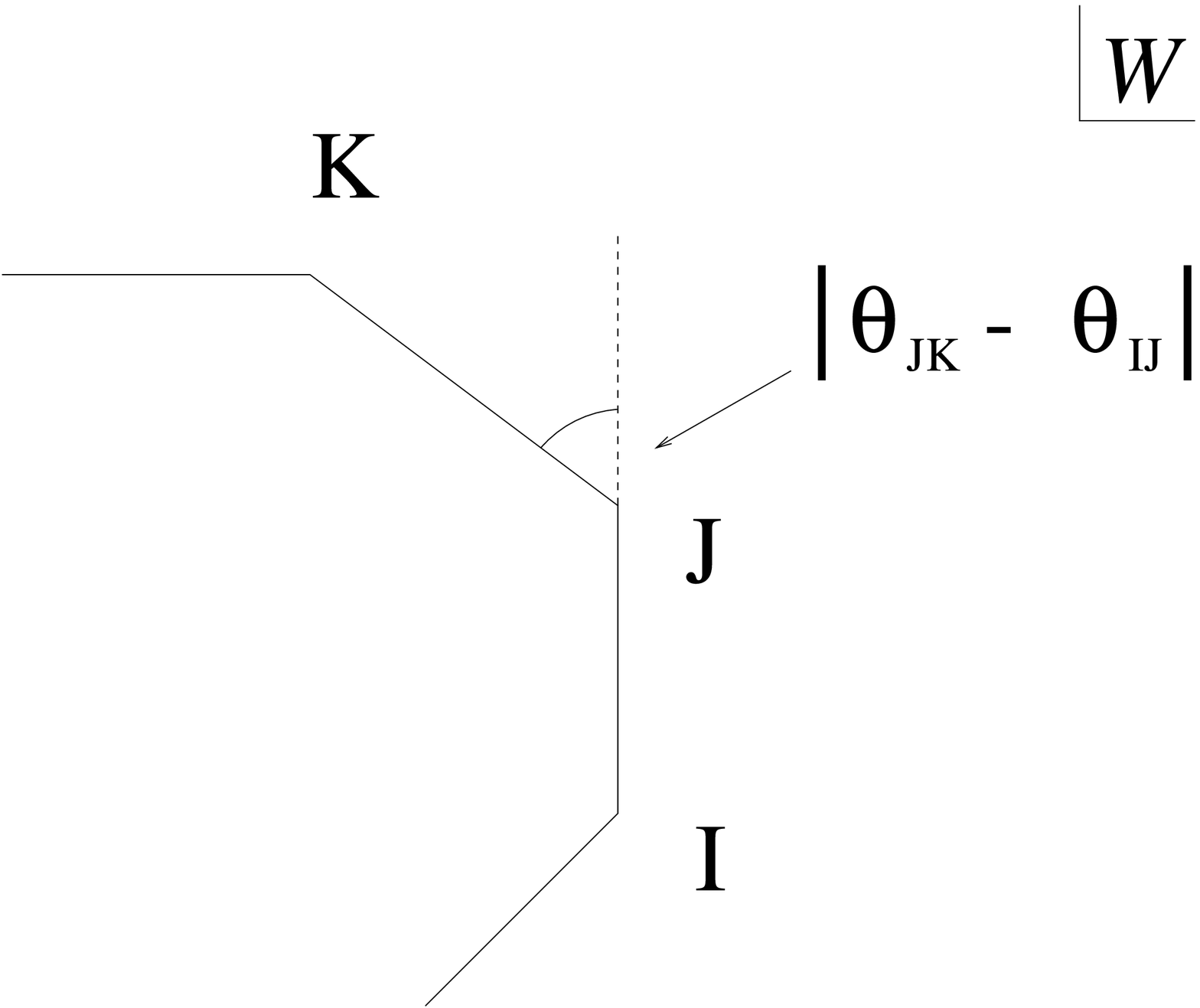} \\ 
\mbox{(a)Walls separating vacua $I, J, K$ in $x^1, x^2$ space} 
\hspace{1cm} 
& 
\mbox{(b)BPS polygon in ${\cal W}$ space} 
\end{array}
\end{eqnarray*}
\caption{Walls in $x^1, x^2$ space and BPS polygon in ${\cal W}$ space}
\label{FIG:polygon}
\end{center}
\end{figure}

BPS domain wall junction is formed when nonparallel BPS walls meet at a 
junction. 
In regions far away from the junction, the configuration approaches to 
isolated walls asymptotically. 
BPS domain wall is a $1/2$ BPS state and conserves two supercharges. 
These two supercharges are given, from Eqs.(\ref{conseved_charge_1st}) 
and (\ref{conseved_charge_2st}), 
in terms of central charges $Z_1$ and $Z_2$ for the wall. 
Let us take a general 
 Wess-Zumino model in Eq.(\ref{generalWZmodel}) and examine if a domain 
wall junction can be formed where $N$ different vacua appear in asymptotic 
regions. 
These $N$ vacua correspond to $N$ points in the 
complex plane of superpotential ${\cal W}$. 
The field configuration of the junction at infinity is mapped to 
a straight line connecting these $N$ vertices \cite{CHT}, \cite{Saffin}. 
In order to have a balance of force, 
this polygon has to be convex \cite{CHT}. 
We set the origin of 
the ${\cal W}$ space at an arbitrary point 
inside this BPS polygon and 
denote the value of the superpotential 
at the  $I$-th vacuum as ${\cal W}_I$, for $I=1,\ldots , N$, 
as illustrated in 
 Fig.~\ref{FIG:polygon}(b). 
Let us take the origin in $x^1, x^2$ space as the junction point of 
 these BPS walls. 
If we denote $\theta_{IJ}$ the angle of the half wall separating two 
vacua, $I$ and $J$, as illustrated in Fig.~\ref{FIG:polygon}(a), 
the central charges $Z_1$ and $Z_2$ of this
wall are given by Eq.(\ref{centralchargeZ}) as 
\begin{eqnarray}
\vec{Z}_{IJ}\equiv (Z_1, Z_2)_{IJ} = 2\, [{\cal W}_J^* - {\cal W}_I^*]
\cdot \vec{\omega}_{IJ}\cdot (\mbox{Area}), \\
\vec{\omega}_{IJ}\equiv (\cos (\theta_{IJ}+\pi/2), \sin (\theta_{IJ}+\pi/2)).
\end{eqnarray}
Thus two supercharges conserved at this wall are
\begin{equation}
  Q_1 + e^{i(-\alpha_{IJ}-\theta_{IJ})}\bar{Q}_{\dot{1}}, \qquad
  Q_2 + e^{i(-\alpha_{IJ}+\theta_{IJ})}\bar{Q}_{\dot{2}},
\label{2-charges}
\end{equation}
where $\alpha_{IJ}=\arg ({\cal W}_J-{\cal W}_I)$. 

BPS domain wall junction is a $1/4$ BPS state and conserves only one
supercharge. 
Let us consider the case of $H=H_{\rm II}$ where 
a linear combination of $Q_2$ and $\bar{Q}_{\dot{2}}$ 
is conserved as shown in Eq.(\ref{conseved_charge_2st}). 
This must be the common conserved supercharge for all the 
walls 
\begin{equation}
 \cdots \, = \, Q_2 + e^{i(-\alpha_{IJ}+\theta_{IJ})}\bar{Q}_{\dot{2}} 
\, = \, Q_2 + e^{i(-\alpha_{JK}+\theta_{JK})}\bar{Q}_{\dot{2}} 
\, = \, \cdots. 
\end{equation}
Then the relative angle of the two neighboring walls must be equal to 
the difference of two phases of the differences $\Delta {\cal W}$ of 
the superpotentials for the two walls 
\begin{equation}
 \cdots,\,\, \theta_{JK}-\theta_{IJ}=\alpha_{JK}-\alpha_{IJ}, \,\,
 \cdots.
\label{angle2}
\end{equation}
Moreover 
the field configuration at infinity should move counterclockwise 
in ${\cal W}$ space, 
as we go around the origin counterclockwise in $x^1, x^2$ space. 

Similarly, a linear combination of $Q_1$ and $\bar{Q}_{\dot{1}}$ 
is the common conserved supercharge in the case of $H=H_{\rm I}$. 
We obtain in this case  
\begin{equation}
 \cdots, \,\, \theta_{JK}-\theta_{IJ}=-(\alpha_{JK}-\alpha_{IJ}), \,\,
 \cdots 
\label{angle1}
\end{equation}
and that 
the field configuration at infinity should move clockwise 
in ${\cal W}$ space, 
as we go around the origin counterclockwise in $x^1, x^2$ space. 

Therefore we find that the BPS equations 
(\ref{Be2})--(\ref{Be2vector}) for the case $H=H_{{\rm II}}$ 
should be used if the phase of the superpotential 
${\cal W}$ increases as we go around the origin counterclockwise 
in $x^1, x^2$ space. 
If the phase of the superpotential 
${\cal W}$ decreases as we go around the origin counterclockwise 
in $x^1, x^2$ space, the other BPS equations (\ref{Be1})--(\ref{Be1vector}) 
 for $H=H_{\rm I}$ should be used. 

Next we discuss the sign of the contribution of the 
central charge $Y_3$ to the mass of the junction configuration. 
We can use the Stokes theorem to obtain an expression for 
the central charge $Y_3$ as a contour integral \cite{HKMN} \cite{CHT} 
\begin{equation}
Y_3 =
\int d x^3 \,
i  
\int d^2 x \,
\left[\partial_1\left(K_{i}
 \partial_2 A^i\right)
-\partial_2\left(K_{i}
 \partial_1 A^i\right)
\right] 
=
\int d x^3 \,
i  
\oint K_{i} d A^i, 
\label{y-area-generic}
\end{equation}
where $K_i \equiv \partial K / \partial A^i$ is a derivative 
of the K\"ahler potential $K$. 
This contour integral 
in the field space should be done as a map from a counterclockwise 
contour in the infinity of $z=x^1+ix^2$ plane. 
Only complex fields can contribute to $Y_3$. 
Let us assume for simplicity that there is only one field which can 
contribute to $Y_3$ as in our exact solution.

Eq.(\ref{y-area-generic}) shows that 
the central charge $Y_3$ becomes negative (positive), 
if the asymptotic counterclockwise contour in $x^1, x^2$ is mapped 
into a counterclockwise (clockwise) contour in field space. 
On the other hand, the sign of the contribution of the central charge 
$Y_3$ to the mass of the junction configurtion is determined by 
the formula 
$H=H_{\rm II} = |\langle i Z_1-Z_2 \rangle| +\langle Y_3 \rangle $, 
or 
$H=H_{{\rm I}}=|\langle -i Z_1-Z_2 \rangle| -\langle Y_3 \rangle $. 
The choice of these mass formulas are in turn deterined by the map 
of the asymptotic counterclockwise contour in $x^1, x^2$ space to a 
 counterclockwise or clockwise contour in the superpotential 
space ${\cal W}$. 
Combining these two observations, we conclude that the contribution of the 
central charge $Y_3$ to the mass of the junction configuration is negative 
if the sign of rotations is the same in field space $A^i$ and in 
superpotential space ${\cal W}$, and positive if the sign of rotations is 
opposite.

The field configuration moves counterclockwise in field space 
in our exact solution in (\ref{exact solution}) and then 
the central charge is negative in this solution. 
Since the exact solution satisfies the BPS equation 
for the case $H=H_{\rm II}$, 
 the central charge contributes to the mass of the junction 
configuration negatively. 
Therefore we should not consider the central charge $Y_3$ alone 
as the physical mass of the junction at the center. 
In the junction configuration,  the junction at the  center 
cannot be separated from the walls. 
We also can find a solution for the other case of $H=H_{\rm I}$ 
in our model. 
The solution is just a configuration obtained by a reflection 
$x^1 \rightarrow -x^1$. 
Then the central charge is positive, but the contribution to 
the mass $H=H_{\rm I}$ becomes again negative.  
In either solution, the rotation in field $T$ space has the same sign 
as the rotation in superpotential ${\cal W}$ space. 
Therefore central charge $Y_3$ contributes negatively to the mass 
of the junction, irrespective of the choice of 
$H=H_{{\rm I}}$ or $H=H_{{\rm II}}$.  

More recently this feature of negative contribution of 
 $Y_3$ to the junction mass is studied from a different viewpoint and 
it is argued that this feature is valid in most situations 
except possibly in contrived models \cite{ShifmanVeldhuis}. 
These models, if they exist, should correspond to the case of 
opposite sign of rotations in ${\cal W}$ space and field space. 
%
%


\section{Energy density and central charges}
\subsection{Charge densities}
Our exact solution is useful to examine how the topological charges 
$Z_k$, $Y_k$ and energy of the domain wall junction are distributed in 
$x^1, x^2$ space. 
We shall study their densities and integrated quantities in finite 
regions in this section. 

\subsubsection{$Y$ charge density}
\begin{figure}[htbp]
  \begin{center}
    \leavevmode
    \epsfxsize=10cm\epsfbox{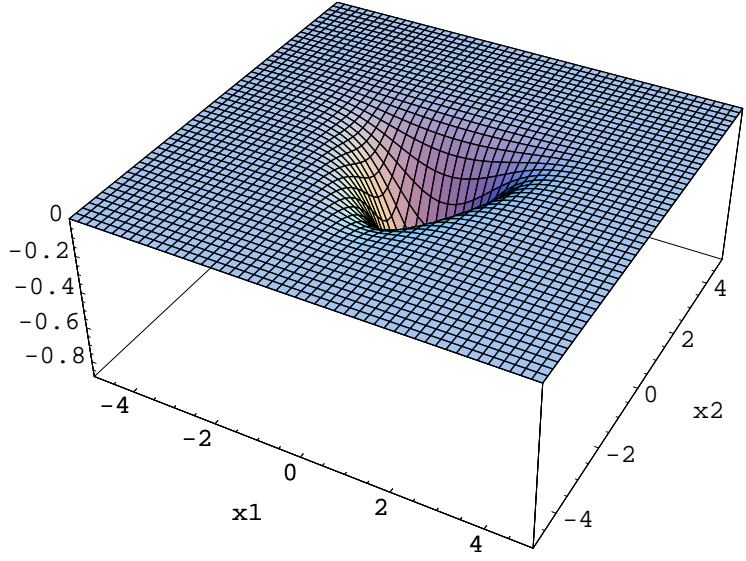}
    \caption{A bird's eye view of ${\cal Y}_3$
.} 
    \label{fig:3d-y3density}
  \end{center}
\end{figure}

The $Y_3$ charge density ${\cal Y}_3=i\epsilon^{3nm}
\partial_n\left(T^*\partial_mT\right)$  is given in our exact solution by 
\begin{eqnarray}
  {\cal Y}_3
&\!\!\! = &\!\!\!
-24\Lambda^4\frac{e^{\sqrt3\Lambda x^2}}{\left[
  e^{\sqrt3\Lambda x^2}+2\cosh \left(\Lambda x^1\right)\right]^3}
\nonumber \\
&\!\!\! = &\!\!\! 
-24\Lambda^4\frac{1}{\left[e^{\frac{2\Lambda r}{\sqrt3}
  \sin\theta}+e^{\frac{2\Lambda r}{\sqrt3}
  \sin\left(\theta+\frac{2\pi}{3}\right)}+
  e^{\frac{2\Lambda r}{\sqrt3}
  \sin\left(\theta-\frac{2\pi}{3}\right)}\right]^3}
\end{eqnarray}
where the cylindrical coordinates $r$ and $\theta$ is used to make 
$Z_3$ symmetry explicit. 
A bird's eye view of the ${\cal Y}_3$ is given in 
Fig.~\ref{fig:3d-y3density}. 
Here and the following, we shall take the unit of 
    $\Lambda\equiv1$ in drawing figures. 
The density is localized near the origin and the $Z_3$ symmetry is 
manifest.

\subsubsection{$Z$ charge density}
We obtain the superpotential as the function of $x^1$ and $x^2$, 
by inserting the solution (\ref{exact solution}) 
\begin{eqnarray}
  {\rm Re}{\cal W}^*&=&-8\Lambda^3\frac{\left(2+3e^{\sqrt3\Lambda x^2}
  \cosh\left(\Lambda x^1\right)+\cosh\left(2\Lambda x^1\right)
  \right)\sinh\left(\Lambda x^1\right)}{\left[
  e^{\sqrt3\Lambda x^2}+2\cosh\left(\Lambda x^1\right)\right]^3}
  \nonumber\\
  {\rm Im}{\cal W}^*&=&2\sqrt3\Lambda^3\frac{e^{\sqrt3\Lambda x^2}
  \left[2+e^{2\sqrt3\Lambda x^2}+6e^{\sqrt3\Lambda x^2}\cosh
  \left(\Lambda x^1\right)\right]}{\left[e^{\sqrt3\Lambda x^2}+
  2\cosh\left(\Lambda x^1\right)\right]^3}.
\label{eq:superpotential}
\end{eqnarray}
The $Z$ charge densities are given by ${\cal Z}_k=2\partial_k{\cal W}^*,
(k=1,2)$ and are found to be 
\begin{eqnarray}
  {\rm Re}{\cal Z}_1&=&-48\Lambda^4\frac{2+e^{2\sqrt3\Lambda x^2}
  \cosh\left(2\Lambda x^1\right)+3e^{\sqrt3\Lambda x^2}
  \cosh\left(\Lambda x^1\right)}{\left[
  e^{\sqrt3\Lambda x^2}+2\cosh \left(\Lambda x^1\right)\right]^4}
  \nonumber\\
  {\rm Im}{\cal Z}_1&=&-{\rm Re}{\cal Z}_2=-48\sqrt3\Lambda^4\frac{e^
  {\sqrt3\Lambda x^2}\sinh\left(\Lambda x^1\right)\left(
  1+2e^{\sqrt3\Lambda x^2}\cosh\left(\Lambda x^1\right)\right)}
  {\left[
  e^{\sqrt3\Lambda x^2}+2\cosh \left(\Lambda x^1\right)\right]^4}
  \nonumber\\
  {\rm Im}{\cal Z}_2&=&48\Lambda^4\frac{e^{\sqrt3\Lambda x^2}\left[
  \cosh\left(\Lambda x^1\right)+e^{\sqrt3\Lambda x^2}\left(
  2+3\cosh\left(2\Lambda x^1\right)\right)\right]}
  {\left[
  e^{\sqrt3\Lambda x^2}+2\cosh\left(\Lambda x^1\right)\right]^4}.
\end{eqnarray}
We can define the effective value of the $Z$ charge which contributes to 
the energy of the junction as $Z_{\rm eff}=-{\rm Re}Z_1+{\rm Im}Z_2$. 
Corresponding effective charge density is given by 
\begin{equation}
  {\cal Z}_{\rm eff}=96\Lambda^4\frac{1+2e^{\sqrt3\Lambda x^2}
  \cosh\left(\Lambda x^1\right)+e^{2\sqrt3\Lambda x^2}\left(
  1+2\cosh\left(2\Lambda x^1\right)\right)}
  {\left[e^{\sqrt3\Lambda x^2}+2\cosh\left(\Lambda x^1\right)\right]^4}.
\end{equation}
Let us note that the effective charge density is $Z_3$ symmetric, 
whereas individual charges ${\cal Z}_1, {\cal Z}_2$ are not. 
%
\subsubsection{Energy density}
\begin{figure}[htbp]
  \begin{center}
    \leavevmode
    \epsfxsize=10cm\epsfbox{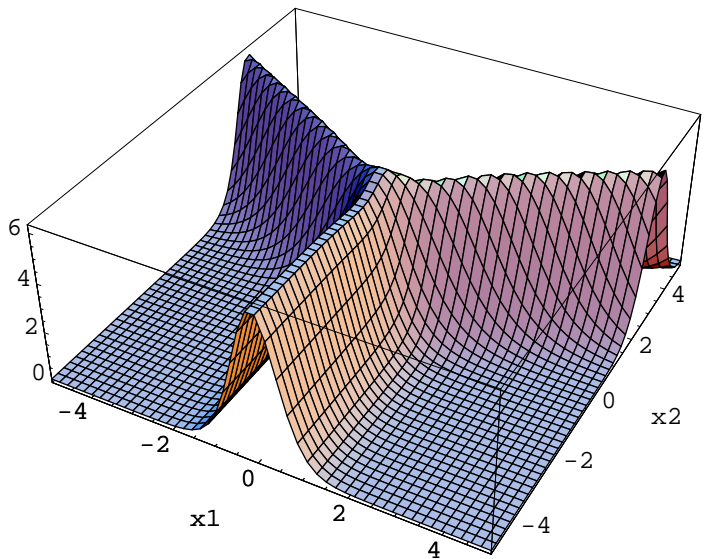}
    \caption{A bird's eye view of the energy density of the junction.} 
    \label{fig:3d-energy-density}
  \end{center}
\end{figure}

Adding ${\cal Z}_{\rm eff}$ and ${\cal Y}_3$ together, 
the energy density of the junction is obtained,
\begin{equation}
  {\cal H}=24\Lambda^4\frac{4+6e^{\sqrt3\Lambda x^2}\cosh
  \left(\Lambda x^1\right)+e^{2\sqrt3\Lambda x^2}
  \left(3+8\cosh\left(2\Lambda x^1\right)\right)}
  {\left[
  e^{\sqrt3\Lambda x^2}+2\cosh \left(\Lambda x^1\right)\right]^4}
\end{equation}
A bird's eye view of ${\cal H}$ is shown in 
Fig.~\ref{fig:3d-energy-density}. 
The energy density is $Z_3$ symmetric as expected.

\begin{figure}[htbp]
  \begin{center}
    \leavevmode
    \epsfxsize=10cm\epsfbox{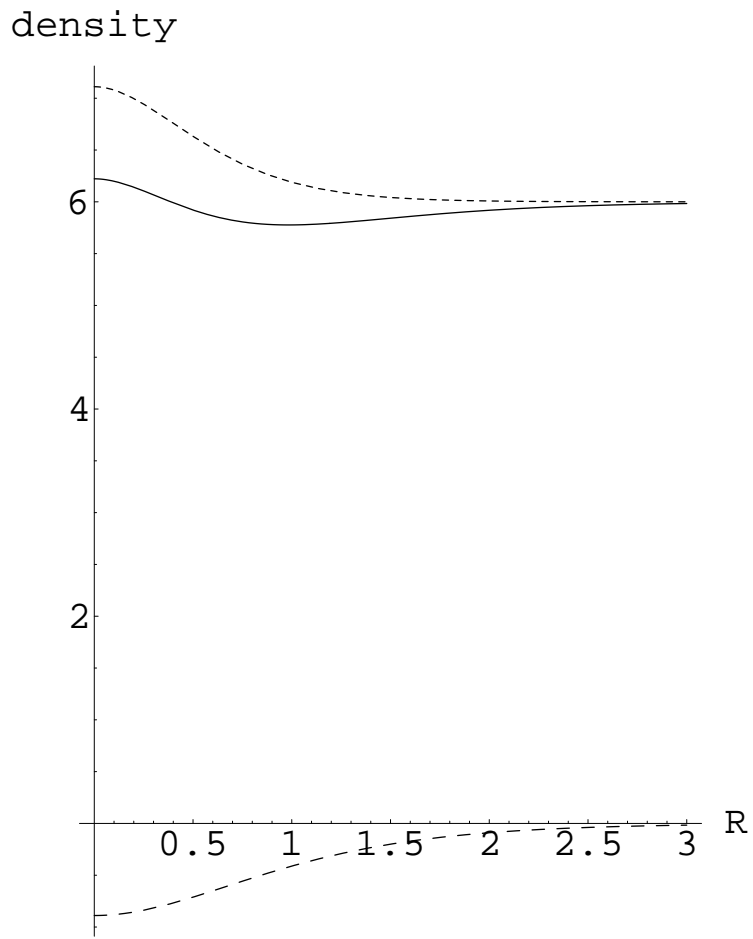}
    \caption{A cross section of the densities 
    ${\cal H}$ (solid line), ${\cal Z}_{\rm eff}$ (short dashed line) and 
    ${\cal Y}_3$ (long dashed line) along the negative $x^2$ direction.} 
    \label{fig:density-comparison}
  \end{center}
\end{figure}

A cross section of the densities, 
${\cal H},{\cal Z}_{\rm eff}$ and 
${\cal Y}_3$ along one of the walls 
(e.g.~negative $x^2$ direction) is shown in Fig.~\ref{fig:density-comparison}.
The $Z_{\rm eff}$ charge contributes to the energy positively
while $Y_3$  does negatively.
Since the decrease of ${\cal Z}_{\rm eff}$ is faster than 
the increase of ${\cal Y}_3$, a small dip is found 
around $R\Lambda\sim1$.
{}Far from the origin there is practically no difference between
${\cal Z}_{\rm eff}$ and ${\cal H}$
because ${\cal Y}_3$ is localized near the origin.
 
\subsection{Charge densities integrated over a region of finite radius}
\begin{figure}[htbp]
  \begin{center}
    \leavevmode
    \epsfxsize=6cm\epsfbox{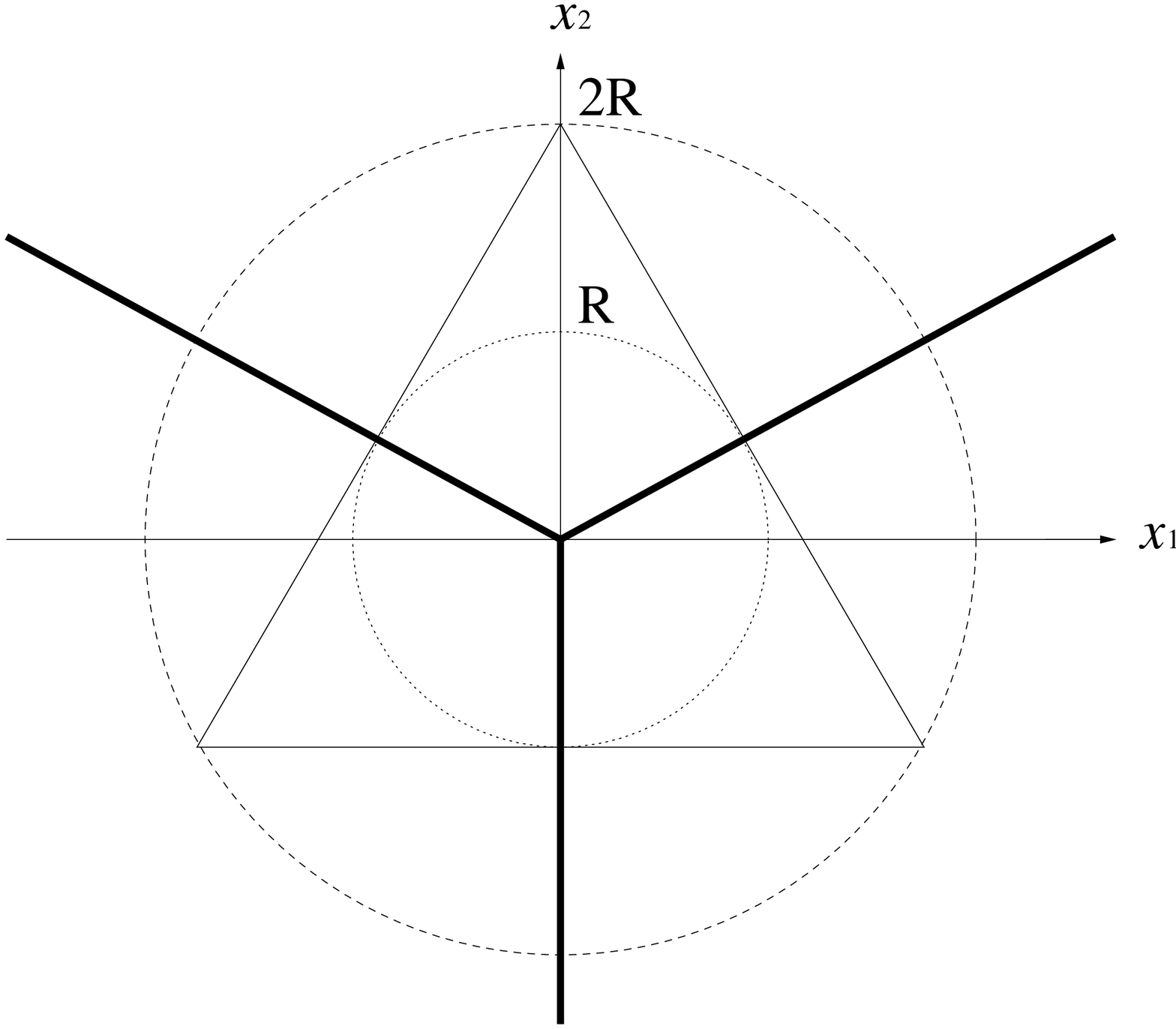}
    \caption{The domains of integration:
    triangle (solid line), inscribed circle (short dashed line) and 
    circumscribed circle (long dashed line).
    The bold lines denote the domain walls forming a 
    junction.} 
    \label{fig:integral-region}
  \end{center}
\end{figure}

In this subsection we shall evaluate 
the central charge densities integrated over a triangular or circular 
region depicted in Fig.~\ref{fig:integral-region}.

\subsubsection{$Y_3$ charge}
Integrating ${\cal Y}_3$ over a triangle whose inscribed 
circle has a radius $R$ as shown in Fig.~\ref{fig:integral-region}, 
we obtain for large $R$ ($R\Lambda\gg1$) 
\begin{equation}
  Y_3^{\rm triangle}(R)=
-2\sqrt3\Lambda^2L_3\left[1-\frac{3\pi}{4}e^{-\sqrt3R\Lambda }
  +{\cal O}\left(e^{-2\sqrt3R\Lambda }\right)\right],
\end{equation}
where $L_3$ denotes the length along the $x^3$ direction.
The leading term agrees with our previous evaluation 
by the step function approximation\cite{HKMN} and the subleading terms
vanish exponentially as $R\rightarrow\infty$.

On the other hand, for  small $R$ ($R\Lambda\ll1$),
we obtain the $Y_3$ charge
\begin{equation}
  Y_3^{\rm triangle}(R)
=-\frac{8}{\sqrt3}\Lambda^2L_3\left[R^2\Lambda^2-R^4
  \Lambda^4-\frac{4\sqrt3}{45}R^5\Lambda^5+{\cal O}
  \left(R^6\Lambda^6\right)\right].
\end{equation}
Notice that the leading term comes from the density at the origin  
multiplied by the area of the circle. 
Although there are no terms of the first or third degree in $R$, 
there is a fifth degree term. 

We can also integrate the ${\cal Y}_3$ over a circle of 
radius $R$.
In this case it is more convenient to use cylindrical coordinates 
($r,\theta$, and $x^3$), 
and the following surface integral formula obtained from the Stokes theorem,
\begin{eqnarray}
 &\!&\! Y_3^{\rm circle}(R)=\int_{-L_3/2}^{L_3/2}dx^3
 \int_0^{2\pi}d\theta\left[iT^{'*}
  \frac{\partial}{\partial \theta}T^{'}\right]\\
  &\!=&\!\frac{8R\Lambda^3L_3}{\sqrt3}\int_0^{2\pi}d\theta
  \frac{\sin\left(\theta-\frac{2\pi}{3}\right)e^{2R\Lambda \cos\theta}+
  \sin\theta e^{2R\Lambda\cos\left(\theta+\frac{2\pi}{3}\right)}+
  \sin\left(\theta+\frac{2\pi}{3}\right)e^{2R\Lambda\cos
  \left(\theta-\frac{2\pi}{3}\right)}}{\left[e^
  {\frac{2R\Lambda}{\sqrt3}\sin\left(\theta+\frac{2\pi}{3}\right)}+
  e^
  {\frac{2R\Lambda}{\sqrt3}\sin\left(\theta-\frac{2\pi}{3}\right)}+
  e^
  {\frac{2R\Lambda}{\sqrt3}\sin\theta}
  \right]^3} \nonumber
\end{eqnarray}
where the $Z_3$ symmetry is manifest.
Expanding the integrand for small $R\ (R\Lambda\ll1)$, 
we obtain
\begin{equation}
 Y_3^{\rm circle}(R)=-\frac{8\pi}{9}\Lambda^2L_3\left[R^2\Lambda^2
  -\frac{1}{2}R^4\Lambda^4+{\cal O}(R^6\Lambda^6)
  \right].
\end{equation}
The leading term is again 
the density at the origin multiplied by the area of the circle.
In contrast to the triangle case, there is no term of odd degree 
in $R$.

\begin{figure}[htbp]
  \begin{center}
    \leavevmode
    \epsfxsize=10cm\epsfbox{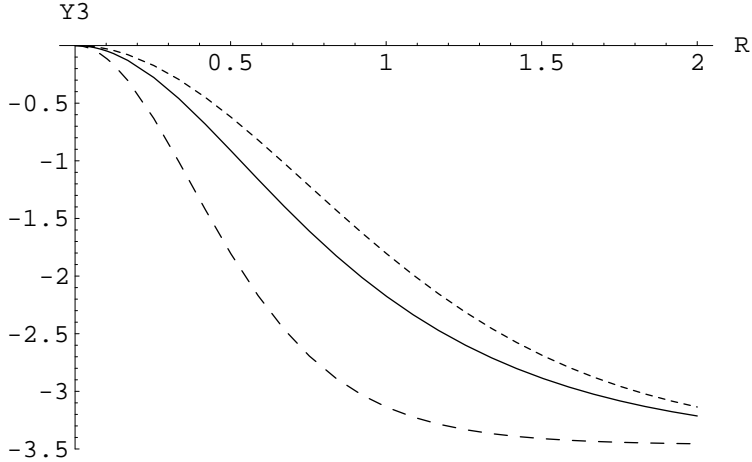}
    \caption{$Y_3(R)$ evaluated for the triangle 
    (solid line), inscribed circle of radius $R$ (short dashed line) 
    and circumscribed circle (long dashed line).} 
    \label{fig:y3-comparison}
  \end{center}
\end{figure}

{}For large $R\ (R\Lambda\gg1)$, 
 we have to perform numerical
integration to evaluate the $Y_3^{\rm circle}(R)$.
We compare the $Y_3(R)$ evaluated for triangle, inscribed and 
circumscribed circle in Fig.~\ref{fig:y3-comparison}.
In the limit of $R\rightarrow\infty$, $Y_3(R)$ for all the regions 
converge to $-2\sqrt3\Lambda^2L_3$ as expected.

\subsubsection{$Z$ charges}
Since the $Z_1$ charge is given by a total derivative 
in $x^1$, we can rewrite the $Z_1$ charge as
\begin{equation}
  Z_1(R)=2\int dx^2\int dx^1\partial_1{\cal W}^*(A^*)
  =2\int_{x^{2-}}^{x^{2+}}dx^2\left[
  {\cal W}^*\left(x^{1+}\left(x^2\right),x^2\right)-
  {\cal W}^*\left(x^{1-}\left(x^2\right),x^2\right)\right],
\end{equation}
where $x^{i \pm}(i=1,2)$ denote the upper and lower bound of the 
domain of integration. 
Since Eq.(\ref{eq:superpotential}) shows that 
${\rm Im}{\cal W}^*(x^1,x^2)$ is 
even and 
${\rm Re}{\cal W}^*(x^1,x^2)$ is odd in $x^1$, 
we obtain 
${\rm Im} Z_1(R)=0$ and ${\rm Re}Z_2(R)=0$ for an 
integration region symmetric in $x^1$ which we shall use. 
Let us note that ${\rm Re}{\cal Z}_1$ (${\rm Im}{\cal Z}_2$) is 
negative (positive) definite.  

{}Firstly we choose as a domain of integration the triangle region 
whose inscribed circle has a radius $R$.
{}For  large $R\ (R\Lambda\gg1)$, we obtain
\begin{eqnarray}
  {\rm Re}Z_1^{\rm triangle}(R)&=&-12\Lambda^2L_3\left[R\Lambda+
  \frac{\sqrt3\pi}{4}e^{-\sqrt3R\Lambda}+{\cal O}
  \left(e^{-2\sqrt3R\Lambda}\right)\right] \nonumber\\
  {\rm Im}Z_2^{\rm triangle}(R)&=&12\Lambda^2L_3\left[R\Lambda+\frac{\sqrt3}{3}
  \left(\frac{\pi}{8}-\frac{4}{3}\right)e^{-\sqrt3R\Lambda}
  +{\cal O}\left(e^{-2\sqrt3R\Lambda}\right)\right].
\end{eqnarray}
The leading linear term represents the contribution of charge density 
per unit length of the wall. 
It is interesting to observe that there are no constant terms. 
The exponentially suppressed terms represent the way the domain wall junction 
configuration converges to isolated walls as $R \rightarrow \infty$. 
The effective value of the $Z$ charge becomes 
\begin{eqnarray}
  Z_{\rm eff}^{\rm triangle}(R)
&=&-{\rm Re}Z_1^{\rm triangle}(R)+{\rm Im}Z_2^{\rm triangle}(R) 
\nonumber\\
&=&24\Lambda^3L_3R+\sqrt3\left(
  \frac{7\pi}{2}-\frac{16}{3}\right)\Lambda^2L_3e^{-\sqrt3R\Lambda}
  +{\cal O}\left(e^{-2\sqrt3R\Lambda}\right).
\end{eqnarray}
{}For small $R\ (R\Lambda\ll1)$, $Z$ charges become 
\begin{eqnarray}
  {\rm Re}Z_1^{\rm triangle}(R)&=&-\frac{32}{\sqrt3}\Lambda^2L_3\left[
  R^2\Lambda^2+\frac{1}{2}R^4\Lambda^4-\frac{2\sqrt3}{9}R^5\Lambda^5
  +{\cal O}\left(R^6\Lambda^6\right)\right]\nonumber\\
  {\rm Im}Z_2^{\rm triangle}(R)&=&\frac{32}{\sqrt3}\Lambda^2L_3\left[
  R^2\Lambda^2-\frac{1}{2}R^4\Lambda^4+\frac{2\sqrt3}{9}R^5\Lambda^5
  +{\cal O}\left(R^6\Lambda^6\right)\right].
\end{eqnarray}
Notice that the leading term represents 
the density at the origin multiplied by the area of the triangle, and that 
there is no term of the first or third degree in $R$.
The effective value of $Z$ charge is  
\begin{equation}
  Z_{\rm eff}^{\rm triangle}(R)=\frac{64}{\sqrt3}R^2\Lambda^4L_3+{\cal O}
  \left(R^6\Lambda^6\right).
\label{eq:z-eff-triangle}
\end{equation}

We can also choose a circle of radius $R$ as a domain of integration.
{}For small $R\ (R\Lambda\ll1)$ we obtain 
\begin{eqnarray}
  {\rm Re}Z_1^{\rm circle}(R)&=&-\frac{32\pi}{9}\Lambda^2L_3\left[
  R^2\Lambda^2-\frac{1}{4}R^4\Lambda^4
  +{\cal O}\left(R^6\Lambda^6\right)\right]\nonumber\\
  {\rm Im}Z_2^{\rm circle}(R)&=&\frac{32\pi}{9}\Lambda^2L_3\left[
  R^2\Lambda^2-\frac{1}{4}R^4\Lambda^4
  +{\cal O}\left(R^6\Lambda^6\right)\right].
\end{eqnarray}
The leading term is again given by the densities 
at the origin multiplied by the area of the circle.
The effective value of the $Z$ charge is 
\begin{equation}
  Z_{\rm eff}^{\rm circle}(R)=\frac{64\pi}{9}\Lambda^2L_3\left[
  R^2\Lambda^2-\frac{1}{4}R^4\Lambda^4
  +{\cal O}\left(R^6\Lambda^6\right)\right].
\end{equation}

\begin{figure}[htbp]
  \begin{center}
    \leavevmode
    \epsfxsize=10cm\epsfbox{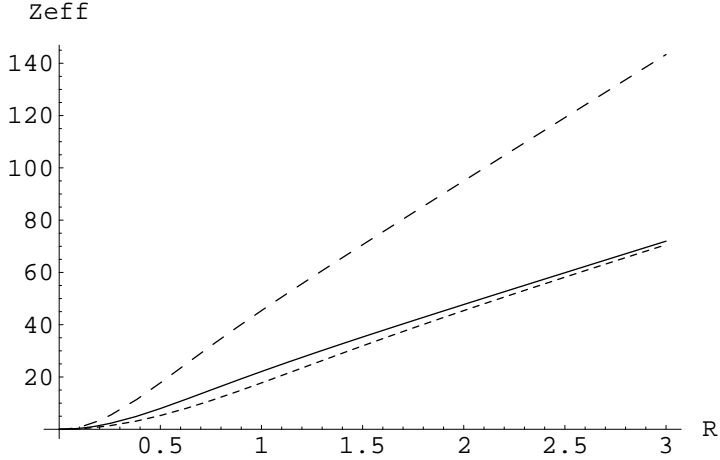}
    \caption{$Z_{\rm eff}(R)$ evaluated for the triangle 
    (solid line),  the inscribed circle 
    of the radius $R$ (short dashed line), and circumscribed circle 
    (long dashed line).} 
    \label{fig:zeff-comparison}
  \end{center}
\end{figure}

A numerical evaluation is needed for large $R$.
We compare the effective $Z$ value evaluated for triangle, inscribed
circle and circumscribed circle in  Fig.~\ref{fig:zeff-comparison}. 
As $R\rightarrow \infty$, the asymptotic slope of 
$Z_{\rm eff}^{\rm ins.circle}(R)$ for inscribed circle 
converges to the same value as that for the triangle. 
In the case of the circumscribed circle, 
 the $Z_{\rm eff}$ 
becomes twice as large as those of the other cases for large $R$, 
since the total length of the walls is twice as long as those of 
the other cases.

\subsubsection{Energy of the junction}
Since our exact solution satisfies the BPS equation corresponding to 
$H=H_{\rm II}$, 
the energy of the junction 
is obtained by adding $Y_3$ and $Z_{\rm eff}$ together 
$H=Z_{\rm eff}+{\rm Y}_3=-{\rm Re}Z_1+{\rm Im}Z_2+{\rm Y}_3$.

{}Firstly we choose the triangle region whose inscribed circle has 
radius $R$.
{}For  large $R\ (R\Lambda\gg1)$, the energy is 
\begin{equation}
  H_{\rm triangle}(R)=
  24\Lambda^3L_3R-2\sqrt3\Lambda^2L_3+\sqrt3\left(
  5\pi-\frac{16}{3}\right)\Lambda^2L_3e^{-\sqrt3R\Lambda}
  +{\cal O}\left(e^{-2\sqrt3R\Lambda}\right).
\end{equation} 
The first linear term can be regarded as the contribution 
from the walls and the second constant term can be regarded as 
the contribution from the junction at the center.
{}For small $R\ (R\Lambda\ll1)$, the energy is
\begin{equation}
  H_{\rm triangle}(R)=\frac{56}{\sqrt3}\Lambda^2L_3\left[
  R^2\Lambda^2+\frac{1}{7}R^4\Lambda^4+\frac{4\sqrt3}{315}R^5\Lambda^5
  +{\cal O}\left(R^6\Lambda^6\right)\right].
\end{equation}
In the case of the circle of radius $R$, 
the energy is given for  small $R\ (R\Lambda\ll1)$ as 
\begin{equation}
  H_{\rm circle}(R)=\frac{56\pi}{9}\Lambda^2L_3\left[R^2\Lambda^2-
  \frac{3}{14}R^4\Lambda^4+{\cal O}\left(R^6\Lambda^6\right)\right].
\end{equation}

\begin{figure}[htbp]
  \begin{center}
    \leavevmode
    \epsfxsize=10cm\epsfbox{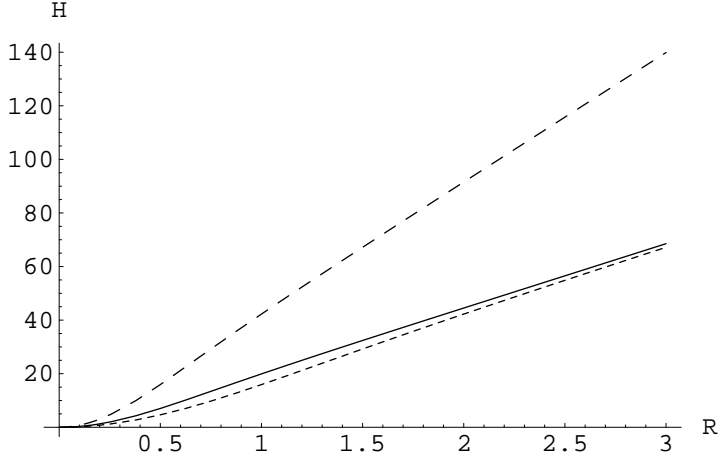}
    \caption{The energy of the junction configuration $H(R)$ evaluated 
    for the triangle (solid),  inscribed circle of radius $R$ 
    (short dashed line) and circumscribed circle (long dashed line).} 
    \label{fig:h-comparison}
  \end{center}
\end{figure}

The energy of the triangle region is compared to those of 
inscribed and circumscribed circles in Fig.~\ref{fig:h-comparison}. 
{}For  large $R\ (R\Lambda\gg1)$, 
the energy $H$ reduces to $Z_{\rm eff}$. 

\begin{figure}[htbp]
  \begin{center}
    \leavevmode
    \begin{eqnarray*}
    \begin{array}{cccc}
    \epsfxsize=8cm\epsfbox{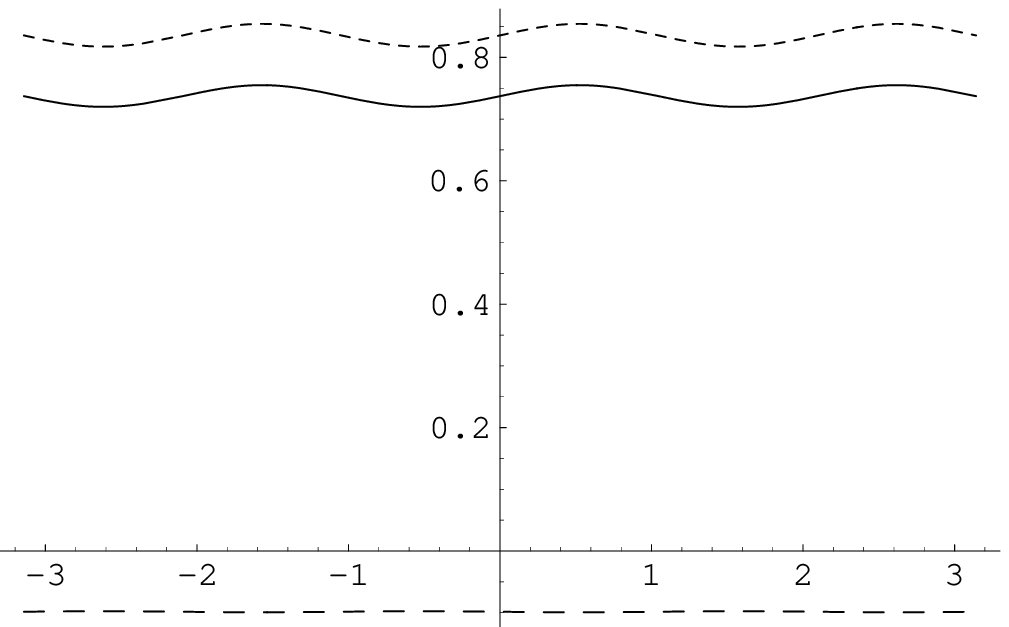} &
    \hspace{1cm}
    \epsfxsize=8cm\epsfbox{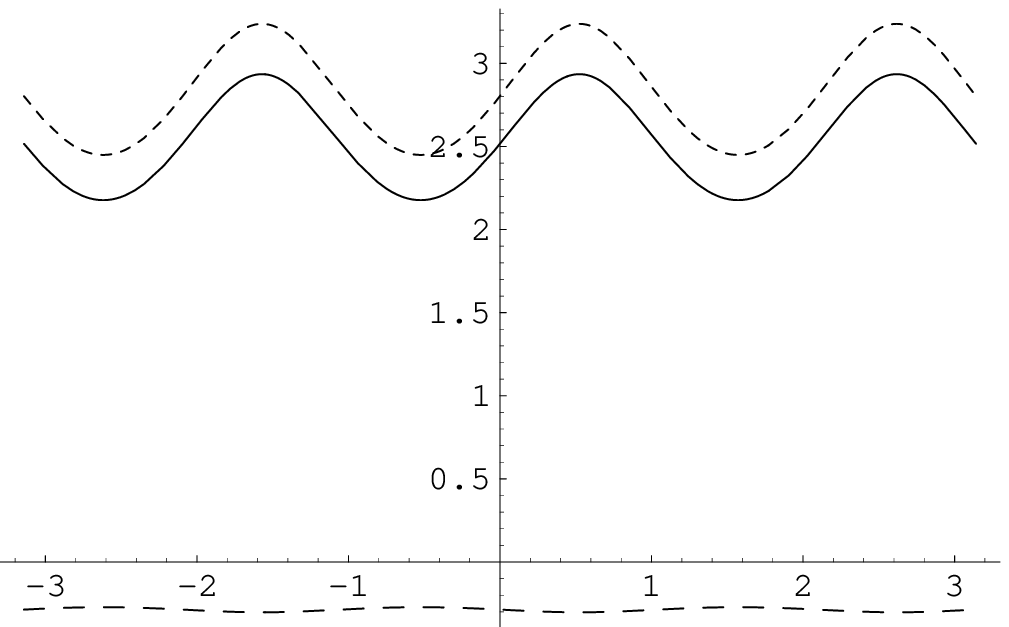} \\
    \mbox{(a) $R\Lambda=0.5$ case} &
    \hspace{2cm}
    \mbox{(b) $R\Lambda=1$ case}   \\
    \epsfxsize=8cm\epsfbox{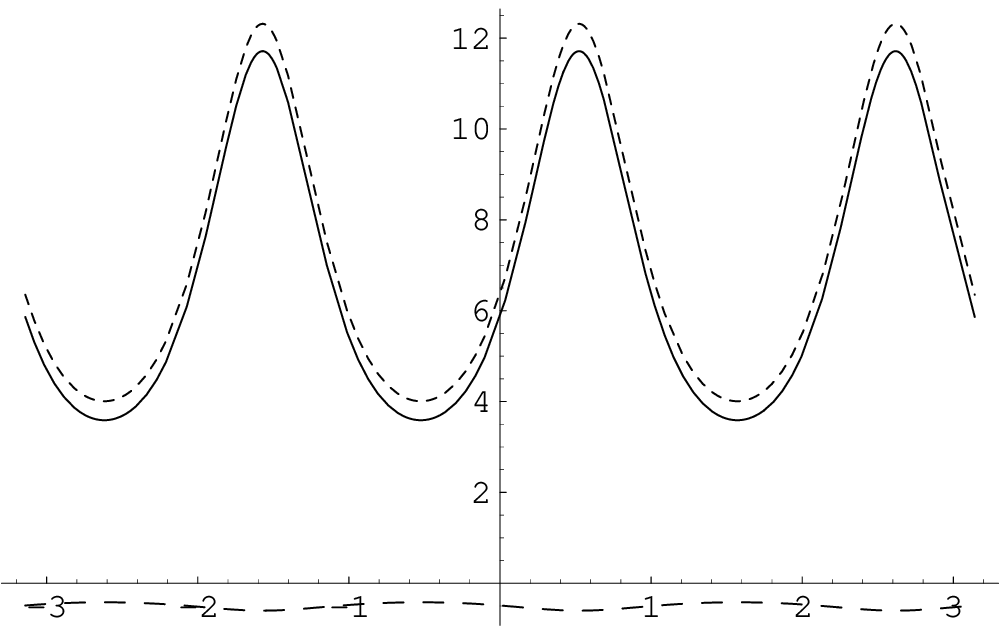} &
    \hspace{1cm}
    \epsfxsize=8cm\epsfbox{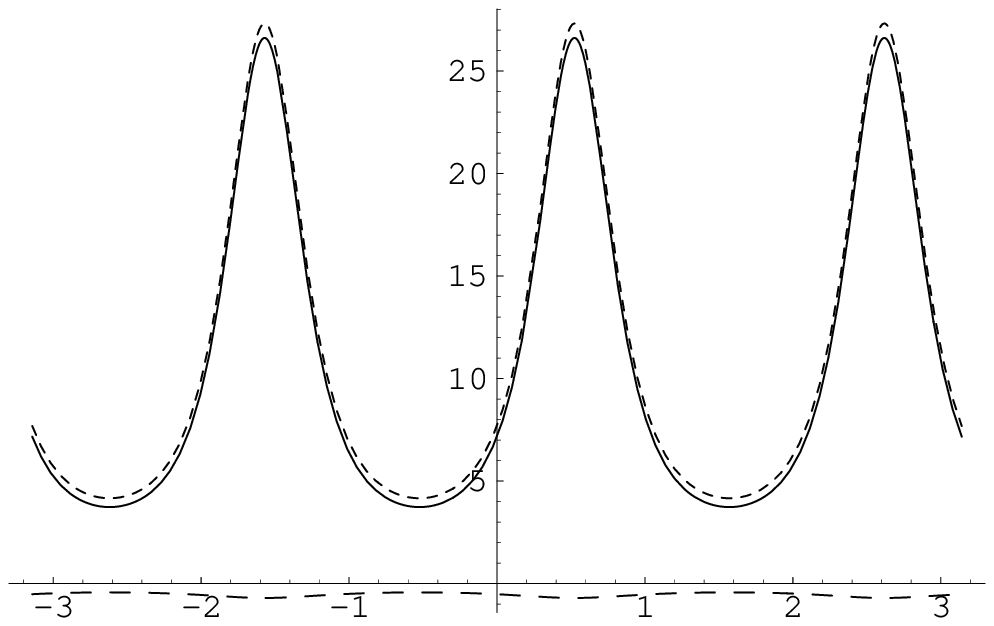} \\
    \mbox{(c) $R\Lambda=2$ case} & 
    \hspace{2cm}
    \mbox{(d) $R\Lambda=3$ case}
    \end{array}
    \end{eqnarray*}
    \caption{The $\theta$--dependence of the energy and charges 
    that are obtained by  
    integrating the densities over the radial direction up to $R$ with 
    $\theta$ fixed.
    In each figure horizontal axis denotes $\theta$,
    and solid, short dashed and long dashed lines correspond to 
    $H$, $Z_{\rm eff}$ and $Y_3$ respectively.} 
    \label{fig:theta-plot}
  \end{center}
\end{figure}

{}Finally we plot the $\theta$--dependence of the energy and 
charges that are obtained by  
integrating the densities from $r=0$ to $r=R$ with 
$\theta$ fixed (see Fig.~\ref{fig:theta-plot}).
In each figure the energy $H$ (solid line) is the sum of the $Z_{\rm eff}$ 
(short dashed line) and $Y_3$ (long dashed line) and $Z_3$ symmetry is 
manifest in their $\theta$--dependence. 
In Fig.~\ref{fig:theta-plot}(a), 
 all the quantities are almost uniform in $\theta$ near the junction 
at the center, 
 reflecting the fact that the junction is a string--like object 
and symmetric around the $x^3$ axis. 
As we move away from the origin, 
 main contribution comes from the direction of the walls 
(in our case $-\pi/2,\pi/6,$ and $5\pi/6$) 
(see Fig.~\ref{fig:theta-plot}(b) and (c)). 
As $R$ grows, $Y_3$ disappears and the energy $H$ approaches 
 $Z_{\rm eff}$ (see Fig.~\ref{fig:theta-plot}(d)).


\renewcommand{\thesubsection}{Acknowledgments}
\subsection{}

One of the authors (H.O.) gratefully acknowledges support from 
the Iwanami Fujukai Foundation.
This work is supported in part by Grant-in-Aid 
for Scientific Research from the Japan Ministry 
of Education, Science and Culture for 
the Priority Area 291 and 707.

\renewcommand{\thesection}{A}
\section{Fermionic contributions to central charges}
\setcounter{equation}{0}
\renewcommand{\theequation}{A.\arabic{equation}}

We shall derive the central charges including fermionic contributions 
in the case of a general Wess-Zumino model 
with an arbitrary superpotential ${\cal W}$. 
For simplicity, K\"ahler metric is assumed to be minimal 
$K_{i j^*} = \delta_{i j^*}$. 
\begin{eqnarray}
{\cal L}&\!=&\! 
-\partial_\mu A^{*j} \partial^\mu A^j + F^{*j}F^j 
+{i \over 2} \partial_\mu \bar \psi^j \bar \sigma^\mu \psi^j 
-{i \over 2} \bar \psi^j \bar \sigma^\mu \partial_\mu \psi^j 
\nonumber \\
&\!+&\! F^j {\partial {\cal W} \over \partial A^j} 
-{1 \over 2} \psi^i \psi^j 
{\partial {\cal W} \over \partial A^i \partial A^j} 
+ F^{*j} {\partial {\cal W}^* \over \partial A^{*j}} 
-{1 \over 2} \bar \psi^i \bar \psi^j 
{\partial {\cal W}^* \over \partial A^{*i} \partial A^{*j}} .
\label{componentWZmodel}
\end{eqnarray}
We have added a surface term to Eq.(\ref{generalWZmodel}) 
to make the variational principle meaningful. 
This is the starting Lagrangian to derive central charges 
and we will not neglect any total divergences from now on. 
The canonical supercurrent is found to be 
\begin{eqnarray}
J^\mu_\alpha 
&\!= &\!
\sqrt2 \left(\sigma^\nu \bar \sigma^\mu \psi^i\right)_\alpha 
\partial_\nu A^{i*} 
+i\sqrt2 \left(\sigma^\mu \bar \psi^i\right)_\alpha F^{i} 
\nonumber \\
&\!= &\!
\sqrt2 \left(\sigma^\nu \bar \sigma^\mu \psi^i\right)_\alpha 
\partial_\nu A^{i*} 
-i\sqrt2 \left(\sigma^\mu \bar \psi^i\right)_\alpha 
{\partial {\cal W}^* \over \partial A^{*i}} \\
\bar J^{\mu \dot \alpha} &\!= &\!
\sqrt2 \left(\bar \sigma^\nu \sigma^\mu \bar \psi^i\right)^{\dot \alpha} 
\partial_\nu A^{i} 
+i\sqrt2 \left(\bar \sigma^\mu \psi^i\right)^{\dot \alpha} F^{i*} 
\nonumber \\
&\!= &\!
\sqrt2 \left(\bar \sigma^\nu \sigma^\mu \bar \psi^i\right)^{\dot \alpha} 
\partial_\nu A^{i} 
-i\sqrt2 \left(\bar \sigma^\mu \psi^i\right)^{\dot \alpha} 
{\partial {\cal W} \over \partial A^{i}} 
\end{eqnarray}
The canonical energy momentum tensor is given by 
\begin{eqnarray}
&
&
T^{\mu\nu}
= 
 \partial^\mu A^j \partial^\nu A^{*j} 
+ \partial^\mu A^{*j} \partial^\nu A^j 
+{i \over 2} \bar \psi^j \bar \sigma^\mu \partial^\nu \psi^j 
+{i \over 2} \psi^j \sigma^\mu \partial^\nu  \bar \psi^j 
\\
&&
+ 
g^{\mu\nu}\left[
-\partial_\lambda A^{*j} \partial^\lambda A^j 
-\left| {\partial {\cal W} \over \partial A^j}\right|^2 
+{i \over 2} \partial_\lambda \bar \psi^j \bar \sigma^\lambda \psi^j 
-{i \over 2} \bar \psi^j \bar \sigma^\lambda \partial_\lambda \psi^j 
-{1 \over 2} \psi^i \psi^j 
{\partial {\cal W} \over \partial A^i \partial A^j} 
-{1 \over 2} \bar \psi^i \bar \psi^j 
{\partial {\cal W}^* \over \partial A^{*i} \partial A^{*j}} 
\right]. \nonumber 
\end{eqnarray}

Canonical quantization gives (anti-) commutation relations 
\begin{equation}
\left[A^i(x), \partial_0A^{*j}(y)\right]_{x^0=y^0}
=
i \delta^3(x-y)\delta^{ij}, 
\qquad 
\left\{\psi^i_\alpha(x), \bar \psi^{j}_{ \dot \beta}(y)\right\}_{x^0=y^0}
=
- \delta^3(x-y)\delta^{ij}\sigma^0_{\alpha \dot \beta} . 
\end{equation}
The anticommutator between supercharges of the same 
chirality gives the supersymmetry algebra 
(\ref{qqantcom}) with the central charge $Z_k$ 
in Eq.(\ref{centralchargeZ}) 
\begin{equation}
Z_k = 2 \int d^3 x \, \partial_k {\cal W}^*(A^*)
\end{equation}
which turns out to have only bosonic contributions. 
The anticommutator between supercharges of the opposite 
chirality gives the supersymmetry algebra 
(\ref{qqantcomy}) with the central charge $Y_k$  
\begin{equation}
Y_k =i \epsilon^{knm}
\int d^3 x \, 
\partial_n \left(A^{*j} \partial_m A^j
-{1 \over 2}\bar \psi^j \bar \sigma_m \psi^j\right), 
\qquad  \epsilon^{123}=1, 
\end{equation}
which has both bosonic and fermionic contributions. 

\renewcommand{\thesection}{B}
\section{Gamma matrices and fermion mode equations}
\setcounter{equation}{0}
\renewcommand{\theequation}{B.\arabic{equation}}

In order to separate variables $(x^1, x^2)$ and $(x^0, x^3)$ for 
spinors, it is most convenient to use a gamma matrix representation 
where the direct product structure of $2 \times 2$ matrices in 
 $(x^1, x^2)$ and $(x^0, x^3)$ becomes manifest. 
One such representation is 
\begin{equation}
\gamma^0 = \rho^1, \quad \gamma^3=i\rho^2, \quad \gamma^1=i\sigma^1\rho^3, 
\quad \gamma^2=i\sigma^2\rho^3, 
\label{eq:standardrep}
\end{equation}
where $\sigma^a$ are Pauli matrices acting on $2 \times 2$ matrices 
and $\rho^a$ acting on indices of blocks of these $2 \times 2$ matrices. 
The four component spinor can be decomposed into a pair of two component 
spinors $\xi$ and $\chi$ in $0+1$ dimensions 
\begin{equation}
 \psi = 
    \left[ 
    \begin{array}{c}       
       \xi \\
       i \chi 
    \end{array}
    \right],
\label{eq:psi}    
\end{equation}
The $B$ matrix for $1+3$ dimensions can be defined as a product 
of $B$ matrices $B^{(1)}$ for $1+1$ dimensions and $B^{(2)}$ 
for $0+2$ dimensions 
\begin{equation}
 B = B^{(1)} B^{(2)}, 
\qquad 
 B^{(1)} = \rho^3, 
\qquad 
 B^{(2)} = -i \sigma^1 .
\end{equation}
The Majorana condition for the $1+3$ dimensional spinor and the 
pseudo-Majorana condition for the $0+2$ dimensional spinor 
are given by 
\begin{equation}
 \psi = B \psi^*, 
\qquad 
 \xi= B^{(2)} \xi^*, 
\quad 
\chi =  B^{(2)} \chi^*,
\label{eq:majoranacond}
\end{equation}
which implies $\xi_1=-i \xi_2^*$ for components of the two 
component spinor $\xi=\left(\xi_1,\xi_2\right)^{T}$, and similarly for $\chi$. 

The Dirac equation for four component fermions in the general 
Wess-Zumino model reads 
\begin{equation}
-i\gamma^\mu \partial_\mu \psi^i 
-\left(
{\partial^2 {\cal W} \over \partial A^{i}\partial A^{j}}
{1 + i \gamma^5 \over 2} + 
{\partial^2 {\cal W}^* \over \partial A^{*i}\partial A^{*j}}
{1 - i \gamma^5 \over 2} 
\right)
\psi^j
 = 0. 
\label{eq:diraceq-4comp}
\end{equation}
The mode equation in $x^1, x^2$ space is defined in terms of two component 
spinors $\xi_n^j$ and $\chi_n^j$ 
\begin{eqnarray}
\left[\delta^{ij}\left(\sigma^1 \partial_1 + \sigma^2 \partial_2 \right) 
-{\partial^2 {\cal W} \over \partial A^{i}\partial A^{j}}
{1 -  \sigma^3 \over 2}  
-{\partial^2 {\cal W}^* \over \partial A^{*i}\partial A^{*j}}
{1 +  \sigma^3 \over 2} 
\right]
\xi^j_n 
    &=&
 - m^{(1)}_n \chi^i_n, 
\nonumber \\
\left[-\delta^{ij}\left(\sigma^1 \partial_1 + \sigma^2 \partial_2 \right) 
-{\partial^2 {\cal W} \over \partial A^{i}\partial A^{j}}
{1 +  \sigma^3 \over 2}  
-{\partial^2 {\cal W}^* \over \partial A^{*i}\partial A^{*j}}
{1 -  \sigma^3 \over 2} 
\right]
\chi^j_n 
    &=&
 - m^{(2)}_n \xi^i_n. 
\end{eqnarray}
The two component spinors $\xi^i_n, \chi^i_n$ satisfy the pseudo-Majorana 
condition (\ref{eq:majoranacond}). 
Then we find that mass eigenvalues are real 
\begin{equation}
m^{(1)}_n = \left(m^{(1)}_n\right)^*, 
\qquad 
m^{(2)}_n = \left(m^{(2)}_n\right)^*. 
\end{equation}
We can make a separation of variables for the 
Dirac equation (\ref{eq:diraceq-4comp}) 
by means of real fermionic fields $c_n$ and $b_n$
\begin{equation}
 \psi^i = \sum_n 
    \left[ 
    \begin{array}{c}       
       c_n(x^0, x^3) \xi_n(x^1, x^2) \\
       i b_n(x^0, x^3) \chi_n(x^1, x^2) 
    \end{array}
    \right]. 
\end{equation}
Using these mode functions we find that the fermion fields 
$(c_n, ib_n)^{T}$ satisfy the Dirac equation in $1+1$ dimensions 
in Eq.(\ref{eq:dirac1+1}). 

This representation can be related to the usual Weyl representation 
in ref. \cite{WessBagger}
by the following unitary matrix $U$ 
\begin{equation}
 U = \frac{1-\sigma^3}{2}\rho^3 + \frac{1+\sigma^3}{2}\rho^2,
 \quad 
 \gamma^\mu_{{\rm Weyl}} = U^\dagger \gamma^\mu U, 
 \quad 
 B_{{\rm Weyl}} = U^\dagger B U^* = \sigma^2 \rho^2. 
\end{equation}
The two component spinor in the Weyl representation is related to 
the components of the four component spinor (\ref{eq:psi}) 
in the representation 
(\ref{eq:standardrep}) in this appendix as 
\begin{equation}
    \left[ 
    \begin{array}{c}       
 \psi_\alpha \\
 \bar \psi^{\dot \alpha}
\end{array}
    \right]_{\rm Weyl}    
=     \left[ 
    \begin{array}{c}       
       \chi_1 \\
       \xi_2 \\
       i \xi_1 \\
       -i \chi_2 
    \end{array}
    \right].    
\end{equation}
Thus we obtain the mode equation (\ref{eq:fermion-mode-weyl1})--
(\ref{eq:fermion-mode-weyl4}) in the Weyl 
representation.


\end{document}